\documentclass[prd,preprint,showkeys,nofootinbib]{revtex4-1}
\usepackage{amsmath}
\usepackage{amssymb}
\usepackage{mathrsfs}
\usepackage{graphicx}

\def\eq#1 { \begin{equation} #1 \end{equation} }
\def\eqn#1{ \begin{eqnarray} #1 \end{eqnarray} }
\def\nn { \nonumber }

\def\half{\frac{1}{2}}

\def\D{\Delta}
\def\l{\lambda}
\def\L{\Lambda}
\def\a{\alpha}
\def\s{\sigma}
\def\eps{\epsilon}

\def\vL{{\vec{L}}}
\def\vM{{\vec{M}}}
\def\vj{{\vec{j}}}

\def\d{\partial}
\def\Int{\mathbb{Z}}
\def\Nat{\mathbb{N}_0}
\def\Reals{\mathbb{R}}
\def\Lint{\mathcal{L}_{\rm int}}
\def\bZ{\overline{Z}}
\def\tZ{\widetilde{Z}}
\def\1PI{{1\rm{PI}}}
\def\V76reg{\overline{{}_7V_6}}
\def\Res{{\rm Res}}
\def\Re{\,{\rm Re}}
\def\Im{\,{\rm Im}}

\def\C#1{ \left\langle #1 \right\rangle }
\def\CL#1{ \left\langle #1 \right\rangle_{\rm L} }
\def\G#1{\Gamma\left(#1\right)}
\def\GG#1{ \Gamma\left[ #1 \right] }
\def \GGG#1#2{\,\Gamma\left[ \begin{array}{l}
      #1 \\
      #2
    \end{array} \right]}
\def\dig#1{\psi\left(#1\right)}
\def\2F1#1#2#3#4{\,\phantom{}_2F_1\left[#1\,,\,#2\,;\,#3\,;\,#4\right] }

\begin{document}
\title{The IR stability of de Sitter: Loop corrections to scalar propagators}

\author{Donald Marolf}
\email{marolf@physics.ucsb.edu}
\author{Ian A. Morrison}
\email{ian\_morrison@physics.ucsb.edu}
\affiliation{University of California at Santa Barbara,
Santa Barbara, CA 93106}
\begin{abstract}
  We compute 1-loop corrections to Lorentz-signature de Sitter-invariant 2-point functions defined by the interacting Euclidean vacuum for massive scalar quantum fields with cubic and quartic interactions.  Our results apply to all masses for which the free Euclidean de Sitter vacuum is well-defined,  including values in both the complimentary and the principal series of $SO(D,1)$. In dimensions where the interactions are renormalizeable we provide absolutely convergent integral representations of the corrections.  These representations suffice to analytically extract the leading behavior of the 2-point functions at large separations and may also be used for numerical computations.  The interacting propagators decay at long distances at least as fast as one would naively expect, suggesting that such interacting de Sitter invariant vacuua are well-defined and are well-behaved in the IR.  In fact, in some cases the interacting propagators decay faster than any free propagator with any value of $M^2> 0$.
\end{abstract}
\maketitle

\section{Introduction}
\label{sec:intro}

While free quantum fields in de Sitter space (dS) have been well understood for some time (see \cite{Allen:1985ux} for scalar fields), interacting de Sitter quantum field theory continues to be a topic of much discussion.  In particular, the literature contains numerous suggestions of possible quantum field theoretic instabilities (see e.g. \cite{EM,TW,polyakov1,Polyakov:2009nq}), many of which have been argued to perhaps lead to decay of the effective cosmological constant.  Our goal here and in \cite{npt} is to address the specific class of such concerns associated with infra-red (IR) divergences of the naive Lorentz-signature de Sitter Feynman diagrams, or more generally those concerns that can be addressed in the context of minimally-coupled scalar fields with $M^2 > 0$.

As we will review, IR divergences arise in generic scalar field theories in Lorentz-signature perturbation theory about the free Hadamard de Sitter-invariant vacuum. (This vacuum is often called the free Euclidean vacuum as it may be defined by analytic continuation from Euclidean signature.)
While such divergences can be avoided at tree level when the fields are sufficiently heavy, they nevertheless arise in loop diagrams.     On the other hand, since Euclidean de Sitter is just a sphere, it is clear that there are no IR divergences in Euclidean signature.  Our goal is to demonstrate that that no pathologies arise from analytic continuation of interacting Euclidean vacuua to Lorentz signature, where they define de Sitter-invariant states.  Specifically we show that, at least through 1-loop order,  the associated Lorentz-signature 2-point functions for massive scalar fields with cubic and quartic interactions are finite and decay at large separations at least as fast as one would naively expect.  This indicates that these Lorentz-signature de Sitter invariant vacuua are both well-defined and well-behaved in the IR.  In particular, it suggests that these vacuua are stable.

Our results apply to all masses for which the free Euclidean de Sitter vacuum is well-defined, i.e. for all $M^2 > 0$,   including values in both the complimentary series and the principal series of $SO(D,1)$. In dimensions where the interactions are renormalizeable, we provide absolutely convergent integral representations of the corrections which allow us analytically extract the leading behavior of the 2-point functions at large timelike separations.   In addition, the representations are amenable to numerical calculations, demonstrating that our methods provide practical tools for calculating Lorentz-signature correlation
functions. We provide a number of checks on our results,  including consistency with known flat-space limits.  The complications associated with both higher loops and higher $n$-point functions will be addressed in \cite{npt}, with similar conclusions. Such results are in qualitative agreement with those obtained using stochastic inflation techniques \cite{SY}, which are expected to be valid in the limit $M \ell \ll 1$, where $\ell$ is the de Sitter length scale.

We begin by briefly reviewing de Sitter field theory in section \ref{sec:prelim}, and by reviewing some useful tools for analytic continuation in section \ref{sec:ACQFT}.  We then compute perturbative corrections to propagators in section \ref{sec:correct} and establish their IR properties, though some details are relegated to the appendices.  An interesting feature is the fact that, in some cases, the corrections enhance the fall-off of the propagator at large times by opening what is effectively a decay channel, even when the daughter particles are heavier than the field under consideration.  This corresponds to the fact that particles in de Sitter space can decay to {\it heavier} particles (see e.g. \cite{ON,Bros:2006gs,Bros:2008sq,Bros:2009bz}) due to the lack of a globally timelike Killing vector field (so that there no conserved notion of energy that is positive definite).  When this occurs, the fall-off of the corrected propagator can be faster than that of any free field with $M^2 > 0$.
Section \ref{sec:disc} then closes with a summary and discussion of general stability issues for de Sitter space.

\section{Field theory in de Sitter space}
\label{sec:prelim}

The following brief review of de Sitter scalar field theory provides an opportunity to fix  conventions and to discuss the IR divergences of naive de Sitter Feynman diagrams.  Consider the $D$-dimensional de Sitter space $dS_D$ for which the metric in global coordinates is

\eq{ \label{eq:metric_dS}
  ds^2 = \ell^2 \left[ - dt^2 + (\cosh t)^2 d\Omega_d^2 \right] ,
}
where $d\Omega_d^2$ is the metric on the unit $d= D-1$ dimensional sphere $S^d$.  Free scalar fields obey the Klein-Gordon equation
\eq{ \label{eq:KG}
  \Box_x \phi(x) = M^2 \phi(x),
}
and define representations of the (connected) de Sitter group
$SO_0(D,1)$.  It is useful to define the dimensionless mass parameter $\s$ by $-\s(\s+d) := M^2\ell^2$.  Throughout  most of our work, the ambiguity
$\s \to -(\s+d)$ will be a redundancy of our description, and symmetry $\s \to -(\s+d)$ will provide a useful check on our calculations.  However, for the moment choosing the branch
\eq{ \label{eq:sigma}
  \s := - \frac{d}{2} + \left[\frac{d^2}{4} - M^2\ell^2\right]^{1/2} ,
}
the standard de Sitter representations may be classified as follows \cite{Vilenkin91}:
\begin{enumerate}
  \item
    complementary series: $ -d/2 < \s < 0$ ,
  \item
    principal series: $\s = -d/2 + i \rho$, $\rho \in \Reals,\;\rho \ge 0$ ,
  \item
    discrete series: $\s = 0,1,2,\dots$ .
\end{enumerate}
We denote the line $\Re \ \sigma = -d/2$ on which the principal series fields live by $\Gamma_P$ below.

\begin{figure}
  \label{fig:sigma}
  \includegraphics{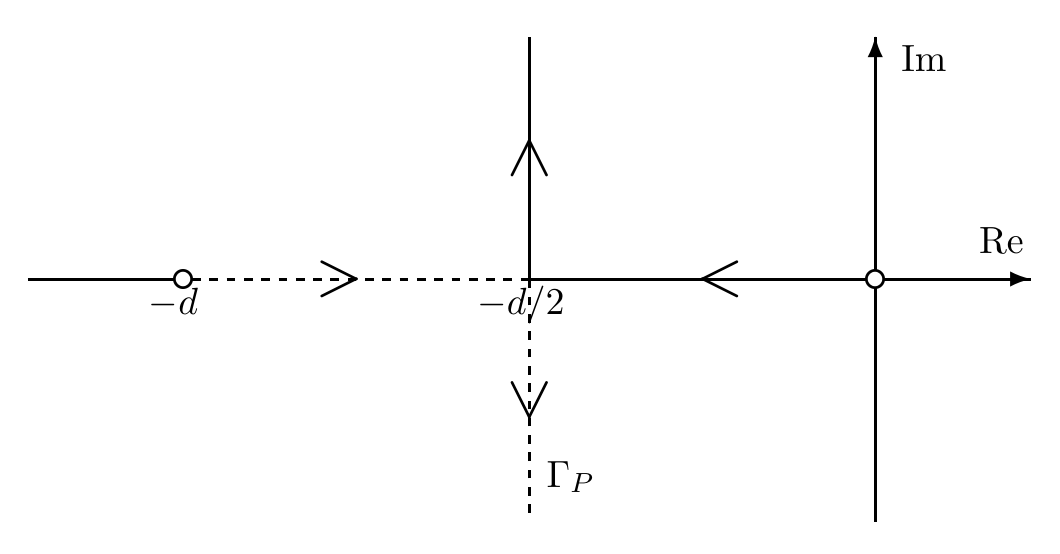}
  \caption{On-shell values of $\s$ in the complex plane for
    massive scalar fields. The solid line denotes the path of $\s$
    for increasing $M^2$ starting from at $\s=0$ for $M^2 = 0$.
    The dotted line shows the path of $-(\s+d)$ for increasing $M^2$
    starting from $-(\s+d) = -d$ for $M^2 = 0$.
    Relatively light fields with $0 < M^2\ell^2 < d^2/4$
    correspond to values of $\s$ and $-(\s+d)$ on the negative
    real axis and belong to the complementary series. Heavier
    fields with $M^2\ell^2 \ge d^2/4$ correspond to complex values
    of $\s$ and $-(\s+d)$ on the line $\Gamma_P$ and belong to the
    principal series.}
\end{figure}

Fields with $M^2 > 0$ correspond to representations in the complementary
and principal series (see figure~\ref{fig:sigma}). In particular,
sufficiently light fields belong to the complementary series while heavier fields
belong to the principal series. In either case, a de Sitter-invariant Green's function $\Delta^\sigma_{xy}$ (with arguments $x,y$) may be defined by analytic continuation from Euclidean signature (i.e., from $S^D$).  We summarize this construction in section \ref{sec:ACQFT} below, but for now we merely state that in the principal and complimentary series the propagator $\Delta^\sigma$ contains terms that fall off like $e^{\sigma |t|}, e^{-(\sigma +d) |t|}$ when one argument is held fixed and the other is taken to large values of $|t|$.   It is important to note that the fastest such decay occurs in the principal series where $\Re \ \sigma =  \Re [-(\sigma +d )] = - d/2.$   We will not consider massless or tachyonic scalars further, as the corresponding free theories do not admit de Sitter-invariant Green's functions \cite{Allen:1985ux}.

Let us now briefly review the IR diverges that arise in calculating naive Lorentz-signature Feynman diagrams.  Before beginning, we emphasize that we discuss Feynman diagrams for correlation functions.  In particular, following the general point of view common in curved spacetime quantum field theory (see e.g. \cite{WaldQFT}),  we view the theory as being defined by its gauge-invariant correlators, with the possible existence of a de Sitter S-matrix being a secondary issue to be investigated at a later stage.

Feynman diagrams in Lorentz signature involve integrating products of propagators over the relevant spacetime, here $dS_D$.  Despite the above exponential decay of de Sitter propagators, this leads to IR divergences due to the exponential growth of the de Sitter volume element $\sim (\cosh t)^{d}$.  For complimentary series fields with $\sigma$ near zero, even the product of 3 or more propagators decays only very slowly so that the most familiar tree-level diagrams (shown in figure 2) diverge.

\begin{figure}
  \label{fig:34}
  \includegraphics{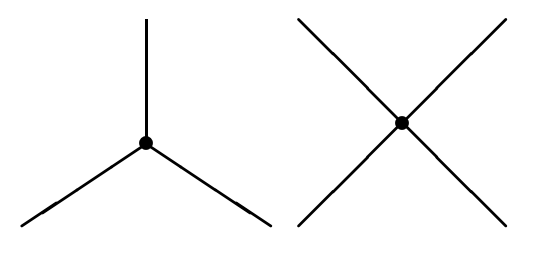}
  \caption{Even the above tree-level diagrams diverge for complimentary
    series fields with $\sigma$ close enough to zero.}
\end{figure}

Furthermore, even in the principal series, IR divergences arise in generic loop diagrams.  Consider for simplicity a correction to the propagator; i.e., a diagram with two external lines.
Let us work in position space and fix the spacetime points $x_1,x_2$ associated with each external line.  Then all vertices must be integrated over $dS_D$.   Consider in particular the integral over the far future region with the relative positions of the vertices held fixed. Then the measure contributes a factor of the de Sitter volume $e^{dt}$ but, since the relative positions of the vertices are held fixed, the integrand is suppressed only by the propagators corresponding to external lines.  From the behaviors quoted above, we see that each contributes a factor of at best $e^{-dt/2}$.  As a result, the integral diverges at least as $\int dt \sim t$.  In particular, so long as both external lines describe fields of the same mass, the integrand contains terms that do not oscillate at large $t$.

It is interesting to note that the above argument applies even to tree-level corrections to propagators; i.e., to any quadratic terms in the Lagrangian (mass terms or kinetic energy terms) which we choose to treat via perturbation theory.  In this context, the above divergences are related to what was termed a failure of the composition principle by Polyakov in \cite{polyakov1} -- see also
\cite{Alvarez:2009kq}.
Of course, despite the divergence of the naive Feynman diagrams, corrections of this form can always be dealt with by simply diagonalizing the quadratic part of the Lagrangian and writing down the resulting free propagators.  At least in this particular case it is clear that there is no problem with the theory itself, but merely with the method of calculation\footnote{As emphasized in
\cite{polyakov1} and further explored in \cite{Higuchi:2009zza},
the failure of naive Lorentz-signature perturbation theory is associated
with the fact that there is no adiabatic theorem in de Sitter space.
Even a slow change of coupling constants in the distant past typically
has finite effects at finite times.  This phenomenon is in turn due to
what is effectively a diverging blueshift due to the rapid contraction
of de Sitter space in the distant past or, what is equivalent, to the
spacelike nature of the past de Sitter boundary $I^-$ (so that geodesics
enter the future light cone of a given point on $I^-$ only at finite times).}.  There is some rough similarity here to the familiar problem of secular divergences in classical mechanics, where an infra-red effect appears to be large due to not properly accounting for finite shifts in frequency\footnote{As noted in \cite{BLHS}, secular divergences are associated even more closely with issues that arise for the special case $M^2=0$ which we do not consider here.}.  The relevant question is whether another method of calculation can remove all IR divergences and define a useful de Sitter-invariant vacuum for the interacting quantum fields.  A natural candidate based on analytic continuation from Euclidean signature is discussed in section \ref{sec:ACQFT} below.

\section{Analytic continuations in de Sitter field theory}
\label{sec:ACQFT}

It is well known that the Euclidean sphere $S^D$ is related to $dS_D$ by analytic continuation.  In particular, the standard metric
\eq{ \label{eq:metric}
  d\Omega_D^2 = \ell^2 \left[d\tau^2 + (\sin\tau)^2 d\Omega_d^2 \right] ,
}
on $S^D$  (\ref{eq:metric_dS}) can be obtained via the Wick rotation
$\tau \to t$ given by
\eq{ \label{eq:AC}
  t = i\left(\tau - \frac{\pi}{2}\right), \quad
  \tau = \frac{\pi}{2} - it .
}
Because $S^D$ is compact, no IR divergences can arise through integrals over $S^D$.  As a result, so long as the linearized field theory admits an $SO(D+1)$ propagator (i.e., so long as there are no massless scalar fields), the familiar Euclidean-signature Feynman diagrams will converge to define an interacting $SO(D+1)$-invariant state on the sphere.  By this we mean that, at each order in perturbation theory, the correlators computed on the sphere will be invariant under $SO(D+1)$ and will satisfy the (Euclidean) Schwinger-Dyson equations.  See e.g. \cite{sphere} for other work involving interacting quantum field theory on $S^D$.

It is therefore natural to attempt to define $SO(D,1)$-invariant states in the Lorentz-signature quantum field theory by analytic continuation from Euclidean signature.  Recall that a state in quantum field theory can be defined by its correlation functions, so it is the correlators upon which the analytic continuation must in fact be performed.  Any set of correlators which satisfies the Schwinger-Dyson equations and appropriate positivity conditions may be considered a valid state of the field theory.  But by the usual arguments the Lorentz-signature Schwinger-Dyson equations are just the analytic continuation of those in Euclidean signature, so the Schwinger-Dyson condition is automatically fulfilled by correlators continued from Euclidean signature.  Similarly, the resulting Lorentz-signature correlators will be invariant under $SO(D,1)$.  Furthermore, in the limit of small couplings, positivity conditions satisfied for free fields cannot be violated by adding perturbative corrections.  In addition, the analogue \cite{Schlingemann:1999mk} for de Sitter
of the Osterwalder-Schr\"ader reconstruction theorem (see e.g. \cite{GJ6}) states that this positivity is guaranteed by reflection-positivity of the Euclidean correlators, which holds at least formally when the potential is bounded below (and which holds rigorously for polynomial potentials bounded below if $D=2$) \cite{GJ10}.

In order to gain more intuition for this procedure, it is useful to describe an alternate (though computationally more difficult) construction of our state.  Because our Lorentz-signature correlators satisfy the Lorentz-signature Schwinger-Dyson equations, they may be thought of as the result of time-evolving initial data from $t=0$.  But at $t=0$ no analytic continuation is required; the Lorentz-signature correlators are precisely the same as that Euclidean correlators up to factors of $i$ associated with explicit time derivatives.  So our Wick-rotated state is identical to what one might call the de Sitter Hartle-Hawking vacuum \cite{Hartle:1976tp} defined by using the Euclidean path integral to compute the state on the $S^3$ at $t=0$ and then evolving away from $t=0$ using the equations of motion.

The existence of perturbative de Sitter-invariant vacuua for interacting (massive) scalar field theories is therefore clear at an abstract level.  In section \ref{sec:correct} below, we demonstrate that there are no hidden subtleties by computing tree and 1-loop corrections to propagators in precisely this way.   Somewhat less trivially, we also explore the large $t$ behavior of the results in order to probe the stability and other IR properties of the resulting de Sitter-invariant states.  Most of our effort will be associated with writing the results in a form appropriate for controlling the analytic continuations at large timelike separations.  The final output will be an integral over functions on $dS_D$ which allows us to read off the large $t$ asymptotics of the Lorentz-signature correlators.  Since the integral converges absolutely, it provides a basis for practical numerical calculations even when it cannot be evaluated exactly.  In the remainder of this section we review two tools that will prove useful in performing the desired analytic continuations.

\subsection{A tool in position space: Embedding distance}
\label{sec:ACED}

Our focus in this work is on loop-corrected 2-point functions.  Since the
vacuum on $S^D$ is invariant under the action of the isometry group $SO(D+1)$,
Euclidean 2-point functions $\C{\phi_i(x_1)\phi_j(x_2)}$ may be written
as functions of the geodesic distance between $x_1$ and $x_2$.
It turns out to be even more convenient to parameterize this separation
by the \emph{embedding distance},
i.e. the length of the chord between $x_1$ and $x_2$ in an ambient
space $\Reals^{D+1}$. This embedding distance may be written in terms
of coordinates on the sphere as
\eq{ \label{eq:Z}
  Z_{12} := Z(x_1, x_2) = \cos\tau_1 \cos\tau_2
  + \sin\tau_1\sin\tau_2 (\vec{x}_1 \cdot \vec{x}_2) ,
}
where $\vec{x}_1$ and $\vec{x}_2$ are unit vectors on the sub-sphere
$S^{D-1}$.
The distance $Z$ is confined to the range $[-1,1]$ with $1(-1)$
the podal(anti-podal) point.

Under the analytic continuation (\ref{eq:AC}) the spherical embedding
distance (\ref{eq:Z}) becomes the $SO(D,1)$-invariant de Sitter embedding
distance
\eq{ \label{eq:Z_dS}
  Z_{12} = -\sinh t_1 \sinh t_2
  + \cosh t_1\cosh t_2 (\vec{x}_1\cdot\vec{x}_2) ,
}
where the embedding space is in this case $\mathbb{M}^{D,1}$.
On $dS_D$, the values of $Z_{12}$ range over all of $\mathbb{R}$;
in particular, the embedding distance satisfies i) $Z_{12} \in [-1,1)$ for
spacelike separations, ii) $Z_{12} = 1$ at coincident points, and
iii) $|Z_{12}| > 1$ for timelike separations.
As a result, a Euclidean correlation function
$\C{\phi_i(x_1)\phi_j(x_2)} = \C{\phi_i\phi_j(Z_{12})}$ may be continued to
the Lorentzian correlator
$\CL{\phi_i(x_1)\phi_j(x_2)} = \CL{\phi_i\phi_j(Z_{12})}$
(or its time-ordered counterpart) simply by continuing $Z_{12}$
from $[-1,1]$ to $\mathbb{R}$.

Of course, one must deal appropriately with branch cuts and singularities for the result to have the desired physical properties.  As may be inferred from the flat-space limit, the correct definition is
\eq{
  \CL{T \phi(x_1) \phi(x_2)} := \C{\phi_i\phi_j(\bZ_{12})} ,
}
where
\eq{ \label{eq:Zbar}
  \bZ_{12} := Z_{12} + i \eps.
}
Similarly, one may define the Lorentz-signature Wightman 2-point function by
\eq{
  \CL{\phi(x_1) \phi(x_2)} :=  \C{\phi_i\phi_j(\tZ_{12})} ,
}
where
\eq{ \label{eq:Ztilde}
  \tZ_{12} = Z_{12} + \bigg\{ \begin{array}{ll}
      + i \eps & \textrm{if $x_1$ is in the future of $x_2$} \\
      - i \eps & \textrm{if $x_1$ is in the past of $x_2$}
    \end{array} .
}

\subsection{A tool in momentum space: Watson-Sommerfeld transformations}
\label{sec:SW}

As in flat space, de Sitter calculations are typically easiest to perform in what is effectively a momentum space representation.  Now, it is well known that there are various subtleties regarding the definition of de Sitter momentum space in Lorentz signature.  For example, the spectrum of the wave operator on $L^2(dS_{D})$ has both continuous and discrete parts \cite{Vilenkin91}.  However, one may avoid all such issues by simply calculating Feynman diagrams in Euclidean signature and using the basis of $L^2(S^D)$ given by the standard spherical harmonics $Y_\vL$ satisfying \cite{Higuchi:1986wu}

\eq{ \label{eq:harmonics1}
  \ell^2 \nabla^2_x Y_\vL(x) = - L(L+d) Y_\vL(x) ,
}
where $\nabla^2$ is the standard scalar Laplacian on $S^D$ with metric (\ref{eq:metric}).
Here $\vL = (L_D,L_{D-1},\dots,L_1)$ is the set of $D$ angular momenta;
the $L_i$ are integers satisfying $L_D \ge L_{D-1} \ge \cdots \ge L_2 \ge |L_1|$.
We will refer to $L_D$ as the total angular momentum. The harmonics
satisfy the orthonormality and completeness relations
\eq{ \label{eq:harmonics2}
  \sum_\vL Y_\vL(x) Y^*_\vL(y) =
  \ell^D \tilde{\delta}(x,y), \quad
  \int_x Y_\vL(x) Y^*_\vM(x) = \ell^D \delta_{\vL\,\vM} .
}
In addition, the harmonics satisfy the following very useful
relation \cite{Drummond}:
\eq{ \label{eq:Y_id}
  \sum_\vj Y_{L\vj}(x)Y_{L\vj}^*(y)
  = \frac{\G{\frac{d}{2}}(2L+d)}{4\pi^{d/2+1}}
  C_L^{d/2}(Z_{xy}) ,
}
where here $\vL = (L,\vj)$, $C_L^\a(x)$ is a Gegenbauer polynomial
(see appendix~\ref{app:Geg}), and $Z_{xy}= Z(x,y)$ denotes (\ref{eq:Z}) with arguments $x,y$.

The usual operations readily express Feynman diagrams on $S^D$ as sums over spherical harmonics or, equivalently, over Gegenbauer polynomials using (\ref{eq:Y_id}).  One might try to obtain useful expressions for de Sitter correlators by analytically continuing such sums over polynomials using (\ref{eq:Zbar}).   However, the $C_L^\a(x)$ are polynomials for integer $L$, so each term in such a sum diverges at large $Z_{xy}$.  But this is precisely the region we want to study, since we wish to determine the behavior of correlators at large timelike separations.

It is therefore useful to rewrite sums of Gegenbauer polynomials $C_L^\a$ as integrals of more general Gegenbauer functions $C_\sigma^\a(x)$ over an appropriate contour $ C$ in the complex plane using a procedure that one might think of as analytic continuation in momentum space.  Specifically, we use a Watson-Sommerfeld transformation (see e.g. \cite{Hartle}): To express a general sum $S = \sum_L s(L)$
as a contour integral in the complex $L$ plane, one first chooses {\it any} function $\tilde s$ such that i) $\tilde s$ agrees $s$ at the values of $L$ appearing in the original sum and ii) $\tilde s$ is analytic in some open neighborhood of the complex $L$-plane around each such point. One then multiplies $\tilde s(L)$ by
a meromorphic kernel $k(L)$ having unit-residue poles at the values of $L$ appearing in the original sum.  One then need only choose an appropriate contour $C_0$ along which to integrate:
\eq{ \label{eq:WSXF}
  S = \sum_L s(L) = \oint_{C_0} \frac{dL}{2\pi i} k(L) \tilde s(L) .
}
Finally, one may attempt to deform the original contour $C_0$ to another contour $C$ over which one has more control.

In our applications, the summand $s(L)$ contains a factor of $C_L^{d/2}(x)$.  We therefore take $\tilde s(L)$ to contain a similar factor in which $C_L^{d/2}(x)$ is a Gegenbauer function (see appendix \ref{app:Geg} for conventions) for general complex $L$. Recall that we wish to evaluate our Feynman diagrams at large $|Z|$.  It is therefore useful to know that for general complex $L$, $C_L^{d/2}(Z)$
is a sum of two terms that behave for large real $Z$ like $Z^L$ and $Z^{-(L + d)}$ (see appendix~\ref{app:Geg}).
As a result, we achieve the most control if we can deform the contour to the line $\Gamma_P$ associated with principal series values of $\sigma$ (i.e., on which $\Re \ L = - d/2$) where both terms decay at large $|Z|$ like $|Z|^{-d/2}$.  Our basic goal\footnote{As we will discuss in section \ref{sec:1loop}, one can obtain even more information about the large $|Z|$ behavior by applying additional tricks, but such embellishments are not needed for the most central results.} is to express all diagrams in terms of integrals over $\Gamma_P$, and to carefully study the extra terms that arise as one deforms the contour from $C_0$ to $\Gamma_P$.  If the integrand decays sufficiently rapidly at large $|L|$, then there is no contribution from infinity.  The Lorentz-signature propagator will then decay at
large values of $|Z|$ if all singularities encountered are sufficiently close to $\Gamma_P$.

It is useful to quickly illustrate this technique by computing the free propagator.  Recall that the free propagator
$\D^\s_{xy}$ on the sphere is the unique solution to the inhomogeneous
Klein-Gordon equation
\eq{ \label{eq:2pt_free_SD}
  - (\nabla^2_x - M^2) \D^\s_{xy}
  = - (\nabla^2_y - M^2) \D^\s_{xy}
  = \delta_{xy} .
}
From the above-mentioned properties of spherical harmonics we
immediately see that $\D^\s_{xy}$ may be written
\footnote{
  Dimensional analysis shows the length dimensions $[\dots]$ of the
  following quantities: $    [M^2] = -2,\quad [\phi] = \frac{2-D}{2},\quad [g_n] = \frac{n(D-2)}{2}-D ,$
  where $g_n$ denotes the coupling constant of $n$-field interactions.
  It follows that $[\D^\s] = 2-D$.
}
\eq{ \label{eq:2pt_free_Y}
  \D^\s_{xy}
  = \ell^{2-D} \sum_\vL \frac{Y_\vL(x) Y_\vL^*(y)}{L(L+d) + M^2 \ell^2}
  = \ell^{2-D} \sum_\vL \frac{1}{\l_{L\s}} Y_\vL(x) Y_\vL^*(y) ,
}
where in the second equality we've defined
\eq{ \label{eq:lambda}
  \l_{L\s} := L(L+d)+ M^2 \ell^2 = (L-\s)(L+\s+d) .
}
The expression (\ref{eq:2pt_free_Y}) provides a spectral representation
of $\D^\s_{xy}$ on the space $(x,y) \in S^D \times S^D$. Other
representations may be found by summing over the angular momenta.
First, by using (\ref{eq:Y_id}) to sum over all but the total
angular momentum one obtains
\eq{ \label{eq:2pt_free_C}
  \D^\s(Z)
  = \ell^{2-D} \frac{\G{\frac{d}{2}}}{4\pi^{d/2+1}}
  \sum_{L=0}^\infty \frac{(2L+d)}{\l_{L\s}} C_L^{d/2}(Z) .
}
This is a spectral representation on the interval $Z \in [-1,1]$.
The fact that $\D^\s(Z)$ depends only on the invariant distance
$Z$ is manifest.  Note that in the form (\ref{eq:2pt_free_C}) one may readily extend the definition of $\D^\s(Z)$ to arbitrary real dimensions $d$, a procedure that will prove useful below and for dimensional regularization of UV divergences.

One now wishes to perform the final sum in (\ref{eq:2pt_free}).  To do so,
we take $\tilde{s}(L)$ to be
\eq{
  \label{eq:s}
  \tilde{s}(L) = \frac{(2L+d)}{\l_{L\s}} e^{-i\pi L} C_L^{d/2}(-Z) ;
}
this is just the summand in (\ref{eq:2pt_free_C}) rewritten slightly
by using the Gegenbauer reflection formula (\ref{eq:Geg_reflection}).
We let
\eq{
\label{eq:k}
  k(L) = \frac{\pi e^{i\pi L}}{\sin(\pi L)} = - e^{i\pi L}\GG{-L,\,1+L},
}
which inserts poles of unit residue at all $L \in \mathbb{Z}$ and write
\eqn{
  \D^\s(Z)
   &=&
   \ell^{2-D} \frac{\G{\frac{d}{2}}}{4\pi^{d/2+1}} \frac{(-1)}{\G{d}} \nn \\
   & & \times
   \oint_{C_1} \frac{dL}{2\pi i} \frac{(2L+d)}{\l_{L\s}} \GG{-L,\,L+d}
   \2F1{-L}{L+d}{\frac{d+1}{2}}{\frac{1+Z}{2}} .
   \label{eq:whatever}
}
Here ${}_2F_1(a,b;c;z)$ is the hypergeometric
function and we use a condensed notation for gamma functions presented in
appendix~\ref{app:gamma}. Since the hypergeometric function is singular at
$Z = 1$, the above procedure should be performed with with $Z < 1$; we will later continue to $|Z| > 1$.

\begin{figure}[t]
  \label{fig:2pt_free}
  \includegraphics{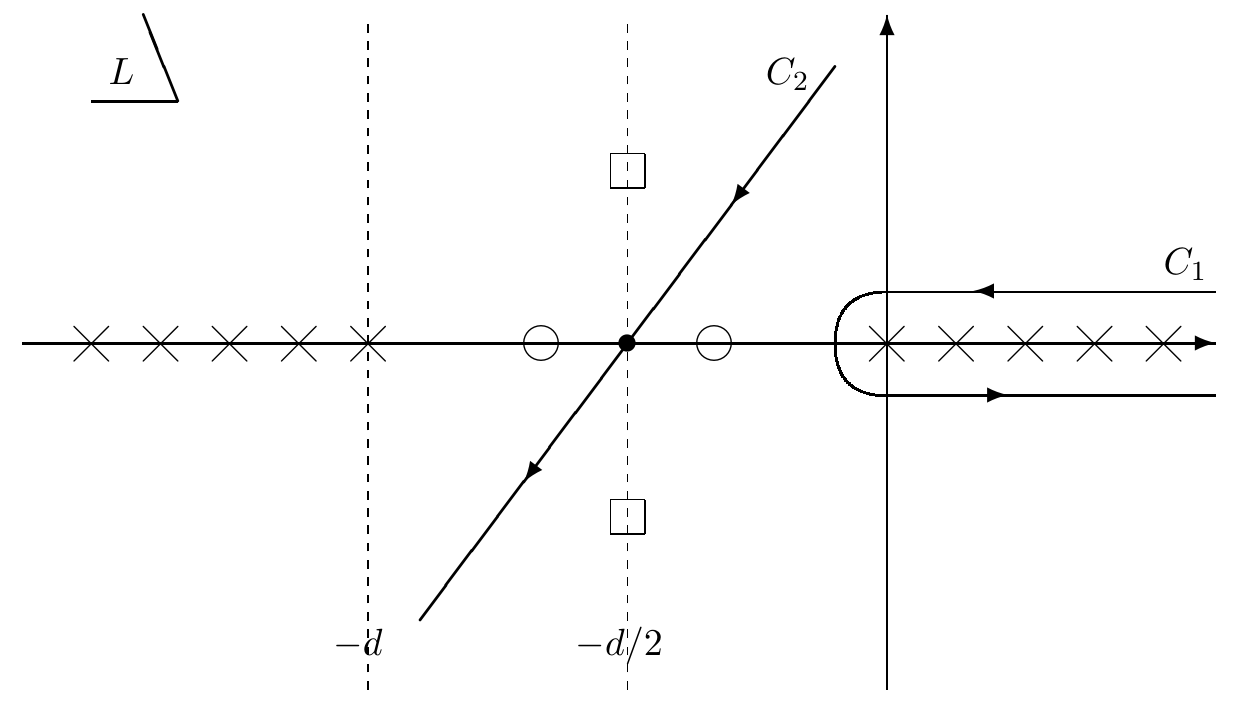}
  \caption{An example of the contour prescription for computing $\D^\s(Z)$.
    The contour $C_2$ is an arbitrary straight line through the
    reflection point $L = -d/2$. Sample $\s$-poles are drawn for
    the principal series (boxes) and complementary series (circles).}
\end{figure}

The contour $C_1$ is depicted in figure~\ref{fig:2pt_free}.
The integrand has poles at $L = 0,1,2,\dots$,
$L = -d, -(d+1), -(d+2), \dots$, and at $L = \s, -(\s+d)$. In fact,
the integrand is antisymmetric under the reflection $L \to -(L + d)$.
Letting $|L| \to \infty$ in the neighborhood of the real axis
the integrand decays as a negative power of $|L|$ -- this is
basically a result of the fact that the original series (\ref{eq:2pt_free_C})
converges -- but, since the factors $e^{\pm i \pi L}$ in (\ref{eq:s}) and
(\ref{eq:k}) cancel in (\ref{eq:whatever}), the integrand decays
exponentially away from the real axis, i.e. at
large $|\Im\,L| \gg 1$ the integrand decays like $e^{- \pi |\Im \,L|}$.
Because of this, we may deform the contour of integration at infinity.
Consider deforming
the contour $C_1$ to the contour $C_2$ defined by a straight line
at any angle to the real axis passing through the reflection point
$L = -d/2$. In the process we deform the contour through exactly one
of the two poles at $L = \s$ or $L = - (\s +d)$, picking
up a residue (see Fig.~\ref{fig:2pt_free}). The remaining line integral
along $C_2$ vanishes due to the antisymmetry of the integrand under
$L \to - (L + d)$. Thus we find that
\eqn{
  \D^\s(Z) &=& \ell^{2-D}
  \frac{\G{\frac{d}{2}}}{4\pi^{d/2+1}} \frac{1}{\G{d}}  \nn \\
  & & \times
  {\rm Res}
  \left\{
    \frac{(2L+d)}{\l_{L\s}} \GG{-L,\,L+d}
    \2F1{-L}{L+d}{\frac{d+1}{2}}{\frac{1+Z}{2}}
  \right\}_{L = \s\;{\rm or}\;-(\s +d)} .
}
The residues at $L = \s$ and $L = -(\s+d)$ are equal (again because the
integrand is antisymmetric under $L \to - (L+d)$), so one readily obtains
\eqn{\label{eq:2pt_free}
  \D^\s(Z) &=& \ell^{2-D} \frac{1}{4\pi^{d/2+1}}
  \GGG{\frac{d}{2},\,-\s,\,\s+d}{d}
  \2F1{-\s}{\s+d}{\frac{d+1}{2}}{\frac{1+Z}{2}}  \\
  &=&- \ell^{2-D} \frac{\G{\frac{d}{2}}}{4\pi^{d/2}\sin(\pi\s)}
  C_\s^{d/2}(-Z),
}
where the definition of the Gegenbauer function (\ref{eq:Geg_def}) was used in the final step.   In this form, the propagator is readily continued to all $Z \in \mathbb{R}$.  In particular, the large $|Z|$ behavior follows from (\ref{eq:Geg_largeZ_branches}) which shows that $\Delta^\s$ is a sum of two terms, respectively proportional to $Z^{\s}$ and $Z^{-(\s + d)}$.

\section{Perturbative Corrections}
\label{sec:correct}

We are now ready to compute perturbative corrections to propagators.  At tree level, such corrections can arise only through interactions of the form $\Lint = g \phi_1(x) \phi_2(x)$, where the fields need not be distinct and $g$ has length units $[g] = -2$.  This interaction can be thought of as resulting from a non-diagonal mass matrix.  At the level of the Lagrangian one may of course use a field redefinition to rewrite this theory in terms of free massive scalar fields, after which one may compute the corrected propagator exactly in either Euclidean or Lorentzian signature.  It is nevertheless useful to understand the perturbative treatment of such terms, not least because they arise as counterterms needed for renormalization.  We therefore briefly discuss such corrections before turning to 1-loop corrections in section \ref{sec:1loop}.

\begin{figure}[t]
  \label{fig:treeLevel}
  \includegraphics{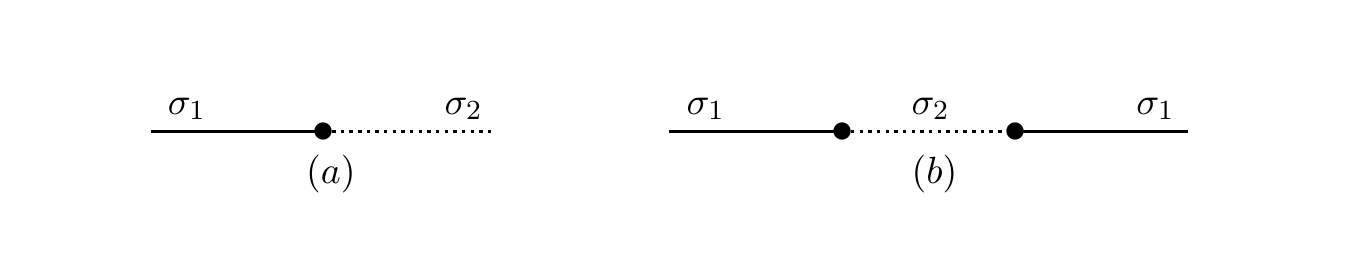}
  \caption{Tree-level contributions to (a) $\C{\phi_1(x_1)\phi_2(x_2)}$
    and (b) $\C{\phi_1(x_1)\phi_1(x_2)}$.}
\end{figure}

On the sphere one may readily compute the leading correction (figure \ref{fig:treeLevel}a):
\eqn{
  \C{ \phi_1(x_1) \phi_2(x_2) }
  &=& - g \int_{y \in S^D} \D^{\s_1}_{1 y} \D^{\s_2}_{y 2}  + O(g)^3 \nn \\
  &=& \frac{g}{M_1^2 - M_2^2}
  \left( \D^{\s_1}(Z_{12}) - \D^{\s_2}(Z_{12}) \right)
  + O(g)^3 .
  \label{eq:treeLevel}
}
The final equality follows from the equation of motion and the fact
that no surface terms arise upon integrating by parts (the latter statement
is \emph{not} true in de Sitter). The degenerate case $M_1^2 = M_2^2$
(e.g., where $\phi_1$ and $\phi_2$ represent the same field) can be
found by taking the limit of (\ref{eq:treeLevel}); the right hand side
becomes $- \d_{M^2} \D^\s(Z_{12}) |_{M^2 = M_1^2}$. Explicit expressions
for $\partial_{\s} C^\alpha_{\s}(Z)$ for half-integer and integer $\a$
may be found in  \cite{Szmytkowski:2007} if desired. The result of course
agrees with what one finds by diagonalizing the quadratic term in the
action, computing the exact propagator, and then expanding the result
perturbatively in $g$.

However, it is useful to note that (\ref{eq:treeLevel}) can also be
obtained following the computational scheme outlined in section \ref{sec:SW}.
One simply uses (\ref{eq:2pt_free_Y}) and the orthogonality of spherical
harmonics to convert the integral over $S^D$ to a sum over $\vec L$,
and then uses (\ref{eq:Y_id}) to write the the result as a sum over
Gegenbauer polynomials.  Finally, one may use a Watson-Sommerfeld
transformation very similar to the one described in section \ref{sec:SW}
to perform the final sum over $L$ and obtain (\ref{eq:treeLevel}). In the
degenerate case the derivative $\d_{\s} \D^\s(Z_{12})$ arises from
evaluating a double pole.

No matter how (\ref{eq:treeLevel}) is obtained, the result is straightforward to continue to Lorentzian de Sitter using (\ref{eq:Zbar}).  Again, the result agrees with the correction found by diagonalizing the quadratic term in the action, computing the exact propagator, and then expanding the result perturbatively in $g$.  This makes it clear that any term of the form (\ref{eq:treeLevel}) is precisely a correction to the scalar field mass matrix, and that the corrected propagators will fall-off at large $Z$ in the manner one would expect.

\subsection{1-Loop contributions}
\label{sec:1loop}

\begin{figure}[t]
  \label{fig:4scalar}
  \includegraphics{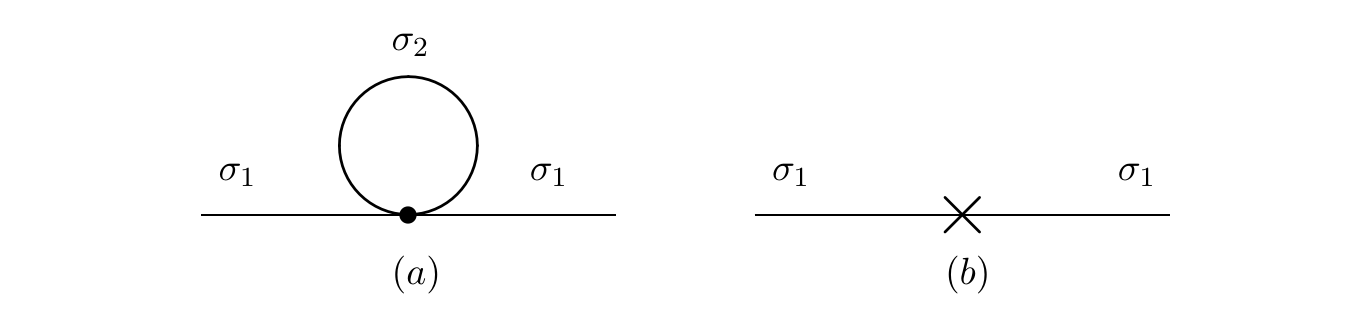}
  \caption{$O(g)$ contributions to $\C{\phi_1(x_1)\phi_1(x_2)}$.
    Diagram (a) is the $1$-loop contribution while (b) is a possible
    counterterm.}
\end{figure}

We now analyze the possible effects of interactions
at $1$-loop order. At this level, only three- and four-particle interactions can contribute to 2-point functions.   For 4-particle interactions of the form $\frac{g}{4} (\phi_1(x))^2(\phi_2(x))^2$ (with $\phi_1$ and $\phi_2$ perhaps representing the same field), the relevant diagrams are those of figure \ref{fig:4scalar}. Both of these diagrams are of the form (\ref{eq:treeLevel}) discussed above.  This is manifest for diagram (b), while it becomes clear for diagram (a) by writing

\begin{equation} - \frac{g}{2} \int_{y \in S^D} \D_{1y}^{\s_1} \D_{yy}^{\s_2} \D^{\s_1}_{y2}
  = - \left[\frac{g}{2} \D^{\s_2}(1)  \right]
  \int_{y \in S^D} \D_{1y}^{\s_1} \D^{\s_1}_{y2} .
   \end{equation}
In other words, diagram (a) of figure \ref{fig:4scalar} is just a constant (given by the propagator at coincident points) times diagram (a) for figure \ref{fig:2pt_free}.  After renormalizing the constant (by using dimensional regularization and perhaps adding the counterterm associated with diagram (b) of figure \ref{fig:4scalar} for the case $m_1 = m_2$),  the result just a (real) correction to the mass of $\phi_1$. Thus the corrected propagators again fall off at large $|Z|$ as one would expect.  One can in fact set the mass corrections to zero by an appropriate choice of renormalization scheme.  Interested readers may find detailed results for the minimal subtraction (MS) scheme  listed in appendix \ref{sec:4p1l} for dimensions $D=3,4$.

We therefore turn directly to the more interesting case of 3-particle interactions, which provide a more computationally difficult example of quantum corrections to the
2-point function.  We consider a theory
with three massive scalar fields and interactions given by
\eqn{ \label{eq:Lint_3P}
  \Lint &=& g \phi_1(x)\phi_2(x)\phi_3(x) \nn \\
  & &
  + \sum_{i=1}^3 \left\{- \half \phi_i(x)
    \left[(\delta\phi_i) \nabla^2_x
      - \left(\delta M_i^2\right) \right]
    \phi_i(x) \right\} .
}
As usual, the results for self-interacting fields can be obtained at the end by taking degenerate limits where one or more masses coincide.

The first term in (\ref{eq:Lint_3P}) provides the 3-particle interaction while the
remaining terms are counterterms which arise from the renormalization
of the fields and bare masses. As for 4-particle interactions,
we can ignore renormalization of the coupling $g$ as it plays
no part in the renormalization of the 2-point function at this level.
The coefficients in (\ref{eq:Lint_3P}) have length units
$\left[ g \right] = \frac{D-6}{2}$,
  $\left[ (\delta \phi_i) \right] = 0$,
  $\left[ \left(\delta M_i^2 \right) \right] = -2 $.
The interaction in (\ref{eq:Lint_3P}) is relevant in
spacetime dimension $D < 6$ and marginal in $D=6$; we will
therefore study this theory in $D=3,4,5,6$.

\begin{figure}[t]
  \label{fig:3P}
  \includegraphics{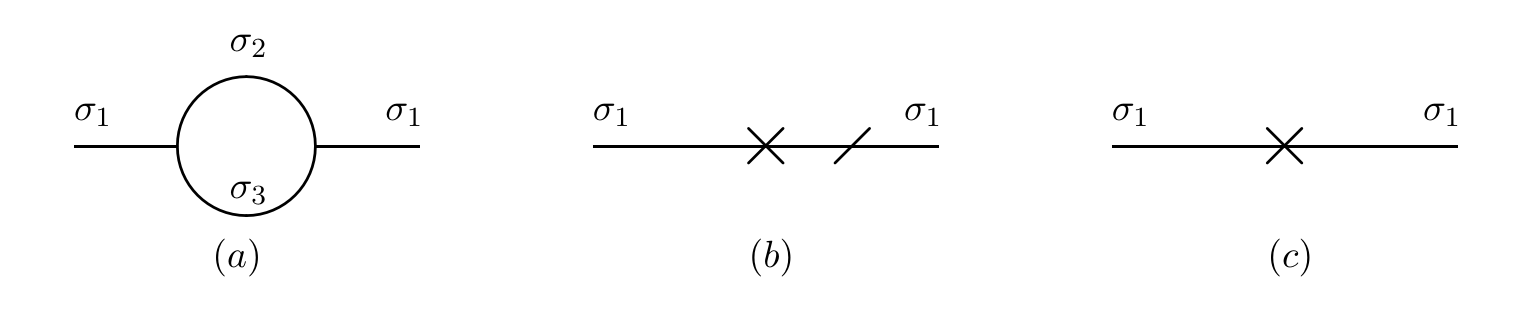}
  \caption{$O(g^2)$ corrections to $\C{\phi_1(x_1)\phi_1(x_2)}$.
    Diagram (a) is the $1$-loop contribution, (b) the counterterm
    due to field renormalization, and (c) the counterterm due
    to mass renormalization. The slash in diagram (b) denotes the
    action of $\nabla^2$.}
\end{figure}

The lowest-order corrections to the scalar 2-point function
$\C{\phi_1(x_1)\phi_1(x_2)}$ occur at $O(g^2)$; they are shown
diagrammatically in figure \ref{fig:3P}. Using the notation of
this figure we write
\eq{
  \C{\phi_1(x_1)\phi_1(x_2)}^{(2)} = (a) + (b) + (c) .
}
Let us first compute diagram $(a)$ on the sphere:
\eq{ \label{eq:a_3P}
  (a) = g^2 \int_{y \in S^D} \int_{y' \in S^D}
  \D^{\s_1}_{1 y} \D^{\s_2}_{y y'} \D^{\s_3}_{y y'} \D^{\s_1}_{y' 2} .
}
To proceed, we expand the product
of two $\D$ distributions on $S^D$ in the basis given by spherical harmonics:
\eqn{ \label{eq:DeltaDeltaRep}
  \D^{\s_1}\D^{\s_2}(Z_{12})
  &=& \ell^{4-2D} \sum_{\vL} \rho_{\s_1 \s_2}(L) Y_\vL(x_1)Y_\vL^*(x_2) \\
  &=& \ell^{4-2D} \frac{\G{\a}}{2\pi^{\a+1}}
  \sum_{L=0}^\infty (L+\a)\rho_{\s_1 \s_2}(L) C_L^{\a}(Z_{12}) .
 }
The spectral function $\rho_{\s_1 \s_2}(L)$ defined by (\ref{eq:DeltaDeltaRep}) will be discussed shortly and is computed in appendix~\ref{app:DeltaDelta}.
As in this appendix, it is convenient to keep track of the spacetime
dimension through the quantity $\a := d/2 = (D-1)/2$.
Inserting (\ref{eq:DeltaDeltaRep}) into (\ref{eq:a_3P}) and using (\ref{eq:2pt_free_Y}), (\ref{eq:harmonics2}), and (\ref{eq:Y_id}) we find
\eq{
  (a) = \ell^{8-2D} g^2 \frac{\G{\a}}{2\pi^{\a+1}} \sum_{L=0}^\infty
  \frac{(L+\a) \rho_{\s_2 \s_3}(L)}{(\l_{L\s_1})^2} C_L^\a(Z_{12}) .
}
The counterterms $(b)$ and $(c)$ are straightforward to compute:
\eqn{
  (b) &=& (\delta\phi_1) \int_{y \in S^D} \D^{\s_1}_{1 y} \Box_y \D^{\s_1}_{y 2}
  = -\ell^{2-D} \frac{\G{\a}}{2\pi^{\a+1}} \sum_{L=0}^\infty
  \frac{(L+a)(\delta\phi_i)L(L+2\a)}{(\l_{L\s_1})^2} C_L^\a(Z_{12}) , \  \\
  (c) &=& - \left(\delta M_1^2\right)
  \int_{y \in S^D} \D^{\s_1}_{1 y}\D^{\s_1}_{y 2}
  = -\ell^{4-D} \frac{\G{\a}}{2\pi^{\a+1}} \sum_{L=0}^\infty
  \frac{(L+\a)\left(\delta M_1^2\right)}{(\l_{L\s_1})^2} C_L^\a(Z_{12}) .
}
Combining our results we have the following expression for
$\C{\phi_1(x_1)\phi_1(x_2)}^{(2)}$:
\eq{ \label{eq:2pt_3P_C}
   \C{\phi_1(x_1)\phi_1(x_2)}^{(2)}
   = \ell^{2-D} \frac{\G{\a}}{2\pi^{\a+1}}
   \sum_{L=0}^\infty f(L) (L+\a) C^\a_L(Z_{12}) ,
}
where we've defined
\eq{ \label{eq:f}
  f(L) := \frac{ g^2\ell^{6-D} \rho_{\s_2 \s_3}(L)
    - \ell^2\left(\delta M^2_1\right)
    - L(L+2\a)(\delta \phi_1)}{(\l_{L \s_1})^2} .
}

Let us now discuss the function $\rho_{\s_1\s_2}(L)$ defined by
the expansion (\ref{eq:DeltaDeltaRep}). From the orthogonality of
Gegenbauer polynomials we may compute
\eq{
  \rho_{\s_1 \s_2}(L) := \ell^{2D-4} \frac{2\pi^{\a+1}}{\G{\a}(L+\a)}
  \frac{1}{A^\a_L} \int_{-1}^{+1} dZ\, (1-Z^2)^{\a-1/2}
  C^\a_L(Z) \D^{\s_1}(Z) \D^{\s_2}(Z) ,
  \label{eq:rhodef}
}
where $A^\a_L$ is the Gegenbauer normalization (\ref{eq:Geg_norm}).
The function $\rho_{\s_1 \s_2}(L)$ is clearly invariant under the actions
\eq{ \label{eq:sigmasyms}
  \s_1 \to -(\s_1 +2\a) , \quad
  \s_2 \to -(\s_2 +2\a) , \quad
  \s_1 \longleftrightarrow \s_2 .
}
Near $Z=1$ the distribution $\D^\s(Z)$ behaves like $\sim (1-Z)^{1/2-\a}$,
so we see that the integral in (\ref{eq:rhodef}) converges for $0 < \a < 3/2$.
We compute $\rho_{\s_1 \s_2}(L)$ for this range of $\a$ in appendix
\ref{app:DeltaDelta}; the result may be written
\eqn{ \label{eq:rho}
  & & \rho_{\s_1\s_2}(L) =
  \Bigg\{ \frac{1}{16\pi^{\a}}
  \frac{\cos(\pi \s_1)}{\sin\pi(\s_1+\a)} \nn \\ & & \times
  \GGG{2-2\a,\,-\s_1,\,L+1,\,2+L-\s_1-\a,\,
    \frac{L-\s_1-\s_2}{2},\,\frac{L-\s_1+\s_2+2\a}{2}}
  {1-\s_1-\a,\,L+\a+1,\,L+1-\s_1,\,\frac{4+L-\s_1-\s_2-4\a}{2},\,
    \frac{4+L-\s_1+\s_2-2\a}{2}} \nn \\ & & \times
  {}_7V_6\left[1+L-\s_1-\a;\,1-\a,\,1-\s_1-2\a,\,1+L,\,
    \frac{L-\s_1-\s_2}{2},\,\frac{L-\s_1+\s_2+2\a}{2}\right] \nn \\ & &
  + (\s_1 \to -(\s_1 + 2\a)) \Bigg\} + (\s_1 \longleftrightarrow \s_2) .
}
Here ${}_7V_6(a;b,c,d,e,f)$ is a so-called very well-poised
${}_7F_6$ hypergeometric function (see appendix~\ref{app:V76}).
The series defining the ${}_7V_6(a;b,c,d,e,f)$ in (\ref{eq:rho})
is absolutely convergent for all complex $L$, $\a$, $\s_1$, and $\s_2$.
We may define $\rho_{\s_1\s_2}(L)$ for complex $\a$ via
the analytic continuation of (\ref{eq:rho}) beyond the interval
$0 < \a < 3/2$. As is discussed in appendix~\ref{app:final}, this
extended $\rho_{\s_1\s_2}(L)$ has poles at $\a = 3/2, 5/2,\dots$. We
also show in this appendix that $\rho_{\s_1\s_2}(L)$
has poles in the complex $L$-plane at
\eqn{ \label{eq:rho_poles}
  L &=& \s_1 + \s_2 - 2n,  \ - \s_1 + \s_2 -2\a - 2n,  \
  \s_1 - \s_2 -2\a - 2n,  \
 - \s_1 - \s_2 -4\a - 2n. \ \ \ \ \ \
}
for $n\in\Nat$.
We will address the meaning of these poles momentarily. An important
property of $\rho_{\s_1\s_2}(L)$ defined in (\ref{eq:rho}) is that
it obeys
\eq{ \label{eq:rho_CC}
  \overline{\rho_{\s_1 \s_2}(L)}
  = \rho_{\overline{\s_1 \s_2}}(\overline{L})
  = \rho_{\s_1 \s_2}(\overline{L})
}
for ``on-shell'' masses $\s_1$ and $\s_2$. The first equality
follows from the fact $\rho_{\s_1 \s_2}(L)$ can be written as an
absolutely convergent series of terms which may be expressed in
terms of Gamma functions, and the Gamma function itself obeys
$\overline{\Gamma(x)} = \Gamma(\overline{x})$. The second equality follows
for on-shell values of $\s_1$ and $\s_2$. ``On-shell'' values of $\s$ are
either (i.) $-\a< \s < 0$, in which case $\s \in \Reals$, or (ii.)
$\s = -\a + i \nu$, $\nu \in \Reals$, for which
$\overline{\s} = -\a - i \nu = - (\s + 2\a)$.

We can now discuss the renormalization coefficients in (\ref{eq:f}).
We use these coefficients to cancel any superficial divergences in
$\rho_{\s_2\s_3}(L)$ and render $f(L)$ finite. For the dimensions
of interest, such superficial divergences occur when $\a = 3/2$ and
$\a = 5/2$~($D = 4$ and $D = 6$). In the
neighborhood $\a = (3-\eps)/2$, $\rho_{\s_2\s_3}(L)$ diverges as
\eq{\label{eq:D4div}
  \rho_{\s_2\s_3}(L)\bigg|_{\a=(3-\eps)/2}
  = \frac{1}{8\pi \eps} + O(\eps^0) .
}
Following the MS scheme, this divergence is cancelled by setting
\eq{
  \left(\delta M_i^2\right)\bigg|_{\a=(3-\eps)/2}
    = \frac{g^2}{8\pi \eps} + O(g^4), \quad
  (\delta \phi_i )\bigg|_{\a=(3-\eps)/2}  = O(g^4) .
}
For $\a = (5-\eps)/2$ we have
\eq{\label{eq:D6div}
  \rho_{\s_2\s_3}(L)\bigg|_{\a=(5-\eps)/2}
  = \frac{-1}{64\pi^3 \eps}\left[ \frac{L(L+5)}{3} + M_2^2\ell^2
    + M_3^2\ell^2-10\right]
  + O(\eps^0).
}
This divergence is cancelled by setting
\eq{
  \left(\delta M_1^2\right)\bigg|_{\a=(5-\eps)/2}
  = - \frac{g^2(M_2^2 + M_3^2)}{64\pi^3 \eps} + O(g^4), \quad
  (\delta \phi_i) \bigg|_{\a =(5-\eps)/2} = - \frac{g^2}{192 \pi^3 \eps}
  + O(g^4) .
}
The expressions for $\left(\delta M_2^2\right)$ and
$\left(\delta M_3^2\right)$ are given by the obvious permutation of
the masses. For $D=3,5$ we set
$\left(\delta M_i^2\right) = (\delta \phi_i) = O(g^4)$.

Let us now return to our expression (\ref{eq:2pt_3P_C}) for
$\C{\phi_1(x_1)\phi_1(x_2)}^{(2)}$. Our task is to rewrite this in a
form well-suited to analytic continuation to Lorentz signature. We proceed in the same way we dealt with the free 2-point function
in (\ref{eq:2pt_free_C}), using a Sommerfeld-Watson transformation defined by the same kernel (\ref{eq:k})
and integrating along a contour $C$ enclosing the poles at
$L =0,1,2,\dots$.
\eq{
  \C{\phi_1(x_1)\phi_1(x_2)}^{(2)}
  = -2 \oint_C \frac{d L}{2\pi i} f(L) (L+\a) \D^{L}(Z_{12}) ,
  \label{eq:2pt_3P_Delta}
}
where we have used (\ref{eq:2pt_free}) to replace Gegenbauer functions by $\D$
distributions for general real $\alpha = d/2$.

\begin{figure}[b]
  \label{fig:3Pcontours}
  \includegraphics{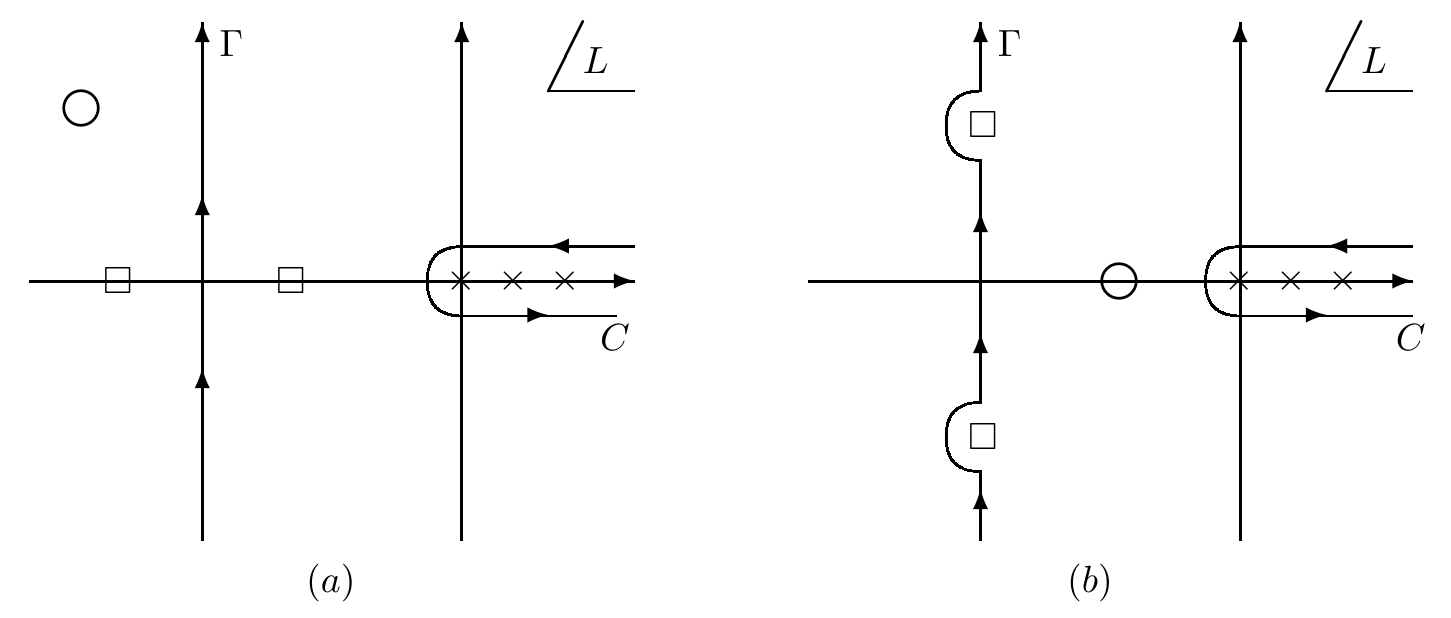}
  \caption{Two examples of the complex $L$-plane. Poles at
    $\s_1$ and $-(\s_1+2\a)$ are marked by $\Box$'s, while the
    pole at $(\s_2 + \s_3)$ is marked by an $\circ$. In example
    $(a)$ $\phi_1(x)$ and $\phi_2(x)$ are in the complementary
    series while $\phi_3(x)$ is in the principal series. Here
    $\Gamma = \Gamma_P$. In
    example $(b)$ $\phi_2(x)$ and $\phi_3(x)$ are in the complementary
    series while $\phi_1(x)$ is in the principal series.  Here $\Gamma$
    differs from $\Gamma_P$ since we take $\Gamma$ to pass to the left of
    poles with $\Re \ L = - \alpha.$ }
\end{figure}

The integrand decays exponentially away from the imaginary
axis like $e^{-\pi |\Im\,L|}$; we can therefore deform the integration contour
away from $C$. We would like to deform the integration contour
to the contour $\Gamma$ along the straight line $\Gamma_P$ defined by $\Re(L) = -\a$ (see figure
\ref{fig:3Pcontours}). By convention, we take $\Gamma$ to pass on the left side of any poles that lie precisely on $\Gamma_P$.

As we deform the contour, we will pick up residues
from any poles we encounter. The integrand in (\ref{eq:2pt_3P_Delta})
has many poles in the $L$-plane. The distribution $\Delta^L(Z)$ has
simple poles at
\eq{
  L = n, \quad
  L = -(n + 2\a), \quad
  {\rm for}\; n \in \Nat .
}
The function $f(L)$ has the simple poles in $\rho_{\s_2\s_3}(L)$ listed
in (\ref{eq:rho_poles}); in addition, the $(\l_{L\s_1})^2$ in the
denominator of $f(L)$ has double-poles at\footnote{
  Furthermore, there are the special cases where
  $\s_1 = -(\s_1+2\a) = -\a$; in this case, due to the $(L+\a)$ in the
  numerator of the integrand, there is a 3rd-order pole at this point.
  There is also the possibility that $\s_1 = \s_2+ \s_3$, in which
  case a double-pole exists at this point. Both of these special cases
  can be found as limiting cases of the more general case so we will
  not treat them explicitly.
}
\eq{
  L = \s_1, \quad L = - (\s_1+2\a) .
}
Despite all these poles, only a very few poles are encountered as we move
the integration contour from $C$ to $\Gamma$.
When $\phi_1(x)$ is in the complementary series then $-\a < \s_1< 0$
and the pole at $L= \s_1$ is on the  right-hand side of $\Gamma$.
When $\phi_1(x)$ is in the principal series both the poles at both
$L = \s_1$ and $L = -(\s_1 +2\a)$ lie on the line $\Gamma_P$. Additionally,
if both $\phi_2(x)$ and $\phi_3(x)$ are in the complementary series
it \emph{may} be that $-\a \le \s_2 + \s_3 < 0$ and even possibly
$-\a \le \s_2 + \s_3-2 < 0$; in these cases the poles at
$L = \s_2+\s_3$ and $L = \s_2+\s_3=2$ lie to the right-hand side of
$\Gamma$. We conclude that
\eqn{
  \C{\phi_1(x_1)\phi_1(x_2)}^{(2)}
  &=& 2 \,\Res\left\{ f(L) (L+\a) \D^{L}(Z_{12}) \right\}_{L=\s_1,
    -(\s_1+2\a)^*,(\s_2+\s_3)^*,(\s_2+\s_3-2)^*}
  \nn \\
  & & + 2 \int_{\Gamma} \frac{dL}{2\pi i} f(L) (L+\a)\D^L(Z_{12}) .
}
Here an asterisk notes that the residue
should only be considered if the pole location has $\Re(L) \ge -\a$.
See figure \ref{fig:3Pcontours} for examples.

Let us first consider the case when $\phi_1(x)$ is a complementary
series field. In this case $-\a < \s_1 < 0$. It is straightforward to
evaluate the residue at $L= \s_1$, which may be written in the useful form
\eqn{
  & & R_1 :=  2\, \Res\left\{ f(L) (L+\a) \D^{L}(Z_{12}) \right\}_{L=\s_1} \nn \\
  &=&
  - \frac{\d}{\d (M^2)} \left\{
    \left[g^2 \,\Re\left\{\ell^{4-D}\rho_{\s_2\s_3}(\s(M^2)) \right\}
      - \left( \delta M^2_1 \right)
      + M^2 (\delta \phi_1) \right] \D^{\s(M^2)}(Z_{12}) \right\}
  \bigg|_{M^2 = M_1^2} .
  \label{eq:s1poleb}
}
Next we examine the integral over $\Gamma$. Inserting $L = -\a + i \nu$
we have
\eqn{ \label{eq:int}
I &:=& 2 \int_{\Gamma_P} \frac{dL}{2\pi i} (L+\a) f(L) \D^L(Z_{12}) \nn \\
  &=& \frac{i}{\pi} \int_{0}^{\infty} d\nu \,
  \nu \left[ f(-\a+i \nu)\D^{-\a+i\nu}(Z_{12}) -
    f(-\a-i \nu)\D^{-\a-i\nu}(Z_{12}) \right] \nn \\
  &=& \frac{i g^2\ell^{6-D}}{\pi} \int_{0}^{\infty} d\nu \,
  \frac{\nu \left[ \rho_{\s_2\s_3}(-\a+i\nu) - \rho_{\s_2\s_3}(-\a-i\nu)
  \right]}{(\l_{-\a+i\nu,\s_1})^2} \D^{-\a+i\nu}(Z_{12}) \nn \\
  &=& - \frac{2 g^2\ell^{2-D}}{\pi} \int_{0}^{\infty} d\nu \,
  \frac{\nu\,\Im\left\{ \rho_{\s_2\s_3}(-\a+i\nu) \right\}}
  {(M^2_1 - M^2_{-\a+i\nu})^2}  \D^{-\a+i\nu}(Z_{12}) .
}
The first equality merely uses the symmetry of the contour under complex conjugation to write the expression as the integral of a quantity that is manifestly real (for real $Z_{12}$). The second equality then follows by using the relations $\l_{-\a- i\nu,\s_1} =  \l_{-\a+ i\nu,\s_1}$ and $\D^{-\a-i\nu}(Z) = \D^{-a+i\nu}(Z)$, the definition of $f(L)$, and the property (\ref{eq:rho_CC}).

Let us now consider the case when $\phi_1(x)$ is in the principal
series. In this case $\s_1 = -\a + i \nu_1$, $\nu_1 \in \Reals$, so
both the poles at $L = \s_1$ and $L = - (\s_1 +2a)$ lie along the
line $\Gamma_P$.  The final result is almost
identical to the result for the case of complementary $\phi_1(x)$. Recall that the contour
$\Gamma$ is indented as shown in figure \ref{fig:3Pcontours} (b).
Thus we have two poles whose residues combine to give the twice the expression on the
right-hand side of (\ref{eq:s1poleb}).  Furthermore,  the integral over $\Gamma$ is of the same form as $I$ (\ref{eq:int}), but with the contour deformed slightly to the left.  It is convenient to remove the indentations in the contour by instead writing the result as the principal part of $I$ added to (\ref{eq:s1poleb}), where the deformation of $\Gamma$ back to $\Gamma_P$ precisely compensates for the extra factor of 2 noted above.

Finally, we must consider the case where both $\phi_2(x)$ and
$\phi_3(x)$ are complementary series fields with sufficiently light
masses such that $-\a < \s_2 +\s_3 < 0$ and possibly $-\a < \s_2 +\s_3-2 < 0$.
In these cases we encounter pole(s) at $L = \s_2 +\s_3$
(and $L = \s_2 +\s_3-2$) as we move the contour. These residues are
easily evaluated using (\ref{eq:rhopoles}):
\eqn{
  \label{eq:last}
  R_2 &:=& 2 \,\Res\left\{ f(L)(L+\a)\D^L(Z_{12})\right\}_{L= \s_2+\s_3}
  \nn \\
  &=& \frac{g^2 \ell^{6-D}}{4\pi^{\a+1}}\frac{1}{(\l_{\s_2+\s_3,\s_1})^2}
  \GGG{-\s_2,\,\s_2+\a,\,-\s_3,\,\s_3+\a}{-\s_2-\s_3,\,\s_2+\s_3+\a}
  \D^{\s_2+\s_3}(Z_{12})  , \\
  \label{eq:lastest}
  R_3 &:=& 2 \,\Res\left\{ f(L)(L+\a)\D^L(Z_{12})\right\}_{L= \s_2+\s_3-2}
  \nn \\
  &=&  \frac{g^2 \ell^{6-D}}{\pi^{\a+1}}
  \frac{1}{(\l_{\s_2+\s_3-2,\s_1})^2}
  \frac{\a (\s_2+\s_3+2\a-2)}{(\s_2+\s_3+\a-1)}
  \GGG{1-\s_2,\s_2+\a-1,1-\s_3,\s_3+\a-1}
  {2-\s_2-\s_3, \s_2+\s_3+\a-2} .\nn \\
}

Assembling our results we have the final expression
\begin{equation}
\label{eq:2pt_3P_final}
   \C{\phi_1(x_1)\phi_1(x_2)}^{(2)} = R_1 + P (I) + R_2 + R_3,
\end{equation}
where $R_1$, $I$, $R_2$, $R_3$ are given respectively in (\ref{eq:s1poleb}), (\ref{eq:int}), (\ref{eq:last}), and (\ref{eq:lastest}).
In (\ref{eq:2pt_3P_final}) the $R_2$ term should be included only
when $-\a < \s_2 + \s_3$ and likewise the $R_3$ term should
only be included when $-\a < \s_2+\s_3-2$. The $P$ in (\ref{eq:2pt_3P_final})
is a reminder to take the principal part in integrating through any pole terms on the axis in the integral $I$.
This result is manifestly real for on-shell masses as it should
be. Earlier we noted in a footnote that that there are two degenerate cases
in which the computation above requires modification, namely
when $\s_1 = -\a$ and when $\s_1 = \s_2+\s_3$. One can find the
correct result for these cases by taking the appropriate limits
of (\ref{eq:2pt_3P_final}).
Finally, the Lorentz-signature correlator $\CL{T \phi_1(x_1)\phi_1(x_2)}^{(2)}$ is defined as
(\ref{eq:2pt_3P_final}) with $Z_{12} \to \tZ_{12}$; likewise, we define
$\CL{\phi_1(x_1)\phi_1(x_2)}^{(2)}$ as (\ref{eq:2pt_3P_final}) with
$Z_{12} \to \bZ_{12}$.

Our final expression (\ref{eq:2pt_3P_final}) is rather complicated.  However, it has two very useful features.  The first is that the remaining integral $I$ converges absolutely for arbitrary $|Z_{12}| > 1$ so long as the contour has been deformed away from any poles in the appropriate manner to compute the principal part. As such, the result is amenable to numerical calculations and gives a practical tool for extracting detailed physics.  The second is that it allows us to extract the large $|Z_{12}| \gg 1$ behavior and so to study the corrected propagator in the deep IR.
At large $|Z_{12}|\gg 1$ the first three terms in (\ref{eq:2pt_3P_final})
have leading behavior, in order,
\eqn{
  |Z_{12}|^{\s_1} \log Z_{12},\quad
  |Z_{12}|^{-\a\pm i \nu},\quad
  |Z_{12}|^{\s_2+\s_3}.
}
The term with the slowest decay provides the leading behavior
at large time-like separation.

As a slight aside we mention that our final expression (\ref{eq:2pt_3P_final})
can easily be brought into the Lehmann-K\"all\'en form of the 2-point
function \cite{Srednicki:2007qs}
\eq{
  \CL{\phi_1(x_1)\phi_1(x_2)}^{(2)}
  = \int_{0}^\infty dM^2 \rho(M^2) \D_{M^2}(Z_{12}) ,
}
where $\rho(M^2)$ (which should not be confused with $\rho_{\s_1\s_2}(L)$)
is the spectral density, and the integral is over positive real
$M^2 > 0$. The integral $I$ in (\ref{eq:int}) is already in the form
of an integral over the principal series masses, so one need only
encorporate the poles $R_1$ (and possibly $R_2$ and $R_3$) into the
integral over $M^2$ by using delta functions in the obvious manner.

It is useful to check our results by taking the flat-space limit of
$\CL{T \phi_1(x_1)\phi_1(x_2)}^{(2)}$.  For convenience, let us suppose that no massless fields arise in this limit.  Thus
all three fields must be in the principal series. For principal
series fields we have only the terms $R_1 + P(I)$.  Now, since
$\D^{\s}(\tZ_{12})$ reduces in this limit to the flat-space propagator $D^{M^2}(x_1-x_2)$, we must have
\eq{
  \ell^{4-D}\rho_{\s_2\s_3}(L) \to \rho^{\rm flat}_{M_2^2 M_3^2}(k^2),
}
where $L^2/\ell^2 \to k^2$ and $\rho^{\rm flat}_{M_2^2 M_3^2}(k^2)$ is the analogous spectral
function of the product of two Minkowski propagators defined by
\eq{
  D_{M_2^2}(x_1-x_2)D_{M_3^2}(x_1-x_2)
  = \int \frac{d^D k}{(2\pi)^D} \rho^{\rm flat}_{M_2^2 M_3^2}(k^2)
  e^{i k (x_1-x_2)} .
}
One
may explicitly check this result for $\a =1$ and $\a=2$ where we have
simplified expressions for $\rho_{\s_2\s_3}(L)$. (We have also explicitly verified that the singular $O(1/\eps)$ terms are equivalent for $\a = 3/2$
and $\a = 5/2$.)  After changing the integration variable in $I$
to $m^2 = \nu^2/\ell^2$, one may take the limit
$\ell \to \infty$ holding fixed $g$, $M_i^2$,
$\left(\delta M_1^2\right)$, and $(\delta \phi_1)$. Noting that  $\Im\{\rho^{\rm flat}_{M_2^2 M_3^2}(-m^2)\} = 0$
for $m^2 < (M_2+M_3)^2$, one finds that the answer agrees with the known flat-space result  \cite{Srednicki:2007qs,Collins:1984xc}.

We conclude this section with a brief discussion of correlator
given by the sum of single particle-irreducible
(1PI) Feynman diagrams. This 1PI correlator may be written
\footnote{
  Implicit in the sum over 1PI diagrams is the assumption
  that $|\Pi(L)/M_1^2| < 1$.
}
\eq{ \label{eq:1PI}
  \C{\phi_1\phi_1(Z_{12})}_{1\rm{PI}}
  =
  \ell^{2-D}\frac{\G{\a}}{2\pi^{\a+1}}
  \sum_{L=0}^\infty \frac{(L+\a)}{\l_{L\s_1}-\Pi(L)} C_L^\a(Z_{12}),
}
where for the diagrams of figure \ref{fig:3P} the dimensionless self-energy is
\eq{ \label{eq:SE}
  \Pi(L) := g^2\ell^{6-2D}\rho_{\s_2\s_3}(L)
  - \ell^2 \left(\delta M_1^2\right)
  - L(L+2\a)\left(\delta \phi_1\right) .
}
This correlator may be analytically continued to de Sitter in
essentially the same way as the $O(g^2)$ correlator above.
An interesting feature we wish to point out is that when
$\phi_1$ belongs to the principal
series and $\Re(\s_2+\s_3) < -\a$, the Lorentz-signature (time-ordered or Wightman) 1PI
correlator decays at large $|Z_{12}|\gg 1$ more rapidly than
\emph{any} free 2-point function.

To see this it is convenient to rearrange the summand of
(\ref{eq:1PI}) slightly before performing the Watson-Sommerfeld
transformation. We use the Gegenbauer recursion relation
(\ref{eq:Geg_recurrence}) to write
\eq{ \label{eq:1PIb}
  \C{\phi_1\phi_1(Z_{12})}_{1\rm{PI}}
  =
  \ell^{2-D}\frac{\G{\a}}{2\pi^{\a+1}}
  \sum_{L=0}^\infty \left\{
    \frac{1}{\l_{L\s_1}-\Pi(L)}
    -
    \frac{1}{\l_{L+2,\s_1}-\Pi(L+2)}
    \right\}  C_L^{\a+1}(Z_{12}) ,
}
where we also use the fact that $C_{-2}^\a(Z)=C_{-1}^\a(Z)=0$ for the
values of $\alpha$ of interest. Using our standard Watson-Sommerfeld
kernel we may write the sum (\ref{eq:1PIb}) as
\eq{ \label{eq:1PIc}
  \C{\phi_1\phi_1(Z_{12})}_{1\rm{PI}}
  = - \ell^{2-D}\frac{\G{\a}}{2\pi^{\a}}
  \int_{\gamma} \frac{dL}{2\pi i}
  \left\{ \frac{1}{\l_{L\s_1}-\Pi(L)}
    -
    \frac{1}{\l_{L+2,\s_1}-\Pi(L+2)}
    \right\}  \frac{C_L^{\a+1}(-Z_{12})}{\sin\pi L} ,
}
where $\gamma$ is a contour parallel to and slightly to the
left of the imaginary axis. Note that at large $|Z_{12}| \gg 1$ the
Gegenbauer function $C_L^{\a+1}(Z_{12})$ behaves like
$(Z_{12})^{L}$ and $(Z_{12})^{-L-2\a-2}$ as compared to $C_L^{\a}(Z_{12})$
which behaves like $(Z_{12})^{L}$ and $(Z_{12})^{-L-2\a}$.
The advantage of changing
the underlying Gegenbauer function from $C_L^{\a}(Z_{12})$ to
$C_L^{\a+1}(Z_{12})$ is that we can shift the contour $\gamma$ as far
to the left as $\Re\,L=-(\a+1)$ while still increasing the decay
of the integrand at large $|Z_{12}|$.

To determine the behavior of the Lorentz-signature correlator at
large $|Z_{12}|\gg 1$ it is sufficient to determine the first pole
encountered as we shift the integration contour $\gamma$ to the left.
For the case of interest the first poles
encountered arise from the first term in brackets and are located
near the on-shell poles $L \approx \s_1$ and $L \approx -\s_1-2\a$.
Letting $\s_1 = -\a + i\nu$ with $\nu \ge 0$ we may write the
on-shell poles in the free theory as $L_\pm := -\a \pm i \nu$.
The location of the on-shell poles in the interacting theory is given by
solving for the zeros of the denominator
\eq{ \label{eq:denom}
  \l_{L\s} - \Pi(L) = L(L+2\a) + M_1^2\ell^2 - \Pi(L) = 0 .
}
Having computed $\Pi(L)$ to $O(g^2)$ we may easily solve
for the $O(g^2)$ corrections to $L_\pm$; the result is
\eq{ \label{eq:OSpoles}
  L_\pm = -\a + \frac{\Im\,\Pi(-\a+i\nu)}{2\nu}
  \pm i \left[ \nu - \frac{\Re\,\Pi(-\a+i\nu)}{2\nu} \right] ,
}
(recall that $\Pi(L)$ is $O(g^2)$).
In writing this expression we've made use of the fact, introduced
above, that $\overline{\rho_{\s_2\s_3}(L)} =\rho_{\s_2\s_3}(\overline{L})$.
These poles contribute residues to the 1PI correlator proportional to
$C_{L_\pm}^{\a+1}(-Z_{12})$; as a result, the leading behavior the
the Lorentz-signature correlators are given by
\eq{ \label{eq:1PIans}
  |Z_{12}|^{-\a+\Im(-\a+i\nu)/2\nu \pm i \omega}
}
with $\omega = \nu - \Re\,Pi(-\a+i\nu)/2\nu$.

The interesting feature of this result is that when
$\Im \Pi(-\a+i\nu) < 0$ the Lorentz-signature correlator decays
\emph{faster} than $|Z_{12}|^{-\a}$, i.e. faster than any free 2-point
function. The only term in $\Pi(-\a +i\nu)$ that can become imaginary
is the function $\rho_{\s_2\s_3}(-\a+i\nu)$. For the cases of $\a = 1$ and
$\a = 2$ (spacetime dimension $D=3$ and $D=5$) one may explicitly
examine the imaginary part of $\rho_{\s_2\s_3}(-\a+i\nu)$ using the
formulas (\ref{eq:rho_D3}) and (\ref{eq:rho_D5})
and find that $\Im(\rho_{\s_2\s_3}(-\a+i\nu)) < 0$.  Furthermore, in the flat-space limit (large $\sigma_1, \sigma_2,\sigma_3$ up to an overall scaling of $\Pi$ to this order) it is known that $\Im\, \Pi \le 0$
\cite{Srednicki:2007qs,Collins:1984xc}.
For $\a = 3/2$ and $\a = 5/2$ ($D = 4$ and $D=6$) with small $\sigma_i$za we
have performed only a small numerical sampling of
$\Im (\rho_{\s_2\s_3}(-\a+i\nu))$, but
in all cases have likewise found that $\Im (\rho_{\s_2\s_3}(-\a+i\nu)) < 0$.

We interpret this result as the appearance of a decay channel that
occurs when $\Re(\s_2+\s_3) < -\a$ and $\phi_1$ is in the
principal series. Note that the requirement $\Re(\s_2+\s_3) < -\a$
does not in general fix the relative size of $M_2^2 + M_3^3$ and  $M_1^2$. Indeed the appearance of this decay channel is
quite generic for any principal series field $\phi_1$; only when
the intermediate states are \emph{very light}, i.e. satisfying
$-\a < \s_2 +\s_3 < 0$, does the behavior of the 1PI correlator
differ from (\ref{eq:1PIans}). The ability of particles to decay
into daughter particles with lighter or heavier masses is
a natural phenomena in de Sitter space
\cite{ON,Bros:2006gs,Bros:2008sq,Bros:2009bz}.
Due to the lack of a globally
time-like Killing vector field,  there is no positive definite conserved energy which would preclude such a process.

\section{Discussion}
\label{sec:disc}

We have computed loop corrections to Lorentz-signature propagators for de Sitter-invariant vacuua in scalar field theories by analytically continuing results from Euclidean signature.  Our results apply to all masses for which the free Euclidean vacuum is well-defined,  including values in both the complimentary series and the principal series of $SO(D,1)$.  We have provided explicit results in dimensions $ D \ge 3$ for which the above  interactions are renormalizeable.   Our results generally take the form of absolutely convergent integral representations sufficient to extract the leading behavior of the Lorentz-signature 2-point functions at large separations.   The absolute convergence implies that such representations are amenable to numerical calculations, demonstrating that our methods provide practical tools for calculating Lorentz-signature correlation functions.  We have provided a number of checks on our results,  including consistency with known flat-space limits.  Our basic methods appear to apply to higher loops as well.

Of course, our use of perturbation theory requires small couplings.  As described recently in \cite{Burgess:2010dd}, perturbative corrections in de Sitter space are controlled by a combination of the coupling and the particle masses which diverges in the limit $M\ell \rightarrow 0$.  In this limit (taking all masses equal), we indeed find that the contributions from the 1-loop diagrams are proportional to $g_4/(M^4\ell^D)$ or $g_3^2/(M^6\ell^D)$, where $g_{3,4}$ are the 3- and 4-point coupling in the Lagrangian.  Our results for the 4-particle interaction agree with those of \cite{Burgess:2010dd} before the application of dynamical renormalization group (DRG) techniques, though it was shown in \cite{Burgess:2010dd} that DRG resummation can ameliorate the $M \rightarrow 0$ growth to some extent. It would be interesting to combine DRG techniques with our Euclidean approach.

With this caveat, we find that the corrected propagators fall off at large separations at least as fast as one would naively expect. Such results are in qualitative agreement with those obtained using stochastic inflation techniques \cite{SY}, which are expected to be valid in the limit $M \ell \ll 1$, where $\ell$ is the de Sitter length scale. Interestingly, for one-loop corrections from 3-particle interactions we found that, in some cases, the corrected (1PI-summed) propagator decays faster than any free propagator with $M^2 > 0$.  This indicates that the vacuum state constructed by analytic continuation of all Euclidean correlators is well-behaved in the IR.  In particular, similar fall-off of higher $n$-point connected correlators would indicate that this state is stable in the following sense: Consider a state $|m\rangle$ constructed by acting on the vacuum with (integrals of) $m$ field operators $\phi_i(x)$ at or near some initial time $t=0$.  Then the $n$-point functions of $|m\rangle$ are just (integrals of) $2m+n$-point functions in our vacuum.  Let us now consider a limit in which the arguments of such an $n$-point function retain fixed relative separations, but in which each argument is taken to some large time in the future; i.e., so that the the $n$-points are far from the $m$ operators originally used to construct the state $|m \rangle$ (which remain near $t=0$).  Decay of connected correlators means precisely that correlators factorize at large separation.
Thus, at large times $t$ any $n$-point function of $|m\rangle$ would approach the product of $\langle m | m \rangle$ with the corresponding $n$-point function in the vacuum.  One might say that, when viewed as a functional on {\it local} products of quantum fields, the large time limit of any above state coincides with our vacuum state.  It is natural to refer to any such vacuum as being stable.  More specifically, when all correlators in a given state factorize in the above limit we will say that the state is an {\it attractor state for local correlators}.

Strictly speaking, a 3-point function or higher is needed to test this notion of stability, while we have computed only propagators here.  We will provide a detailed discussion of higher $n$-point functions elsewhere \cite{npt}, but for now we merely note that our propagator calculations suggest that the perturbative vacuum theory is well-behaved in the IR.

Supposing that the higher $2m+n$-point functions continue to indicate stability of de Sitter-invariant vacuua for (massive) scalar field theories, one may be tempted to ask {\it why} such vacuua should be stable.  For familiar vacuua in flat Minkowski space there is a simple physical answer: each  such vacuum minimizes a positive-definite energy.  But de Sitter space has no positive-definite conserved energy due to its lack of a globally-defined timelike Killing field, so we must search elsewhere for an explanation. The best answer is probably that de Sitter space does admit Killing fields that are timelike in a globally hyperbolic domain known as the ``static patch'' associated with that Killing field.  Such domains may be treated as spacetimes in their own right, with no need to impose extra boundary conditions.  For positive potentials, the associated Hamiltonian is bounded below in this restricted spacetime. As a result, positivity and conservation of this energy forbids instabilities of scalar quantum field theories (with positive potentials) in any static patch.  Any possible instability of de Sitter space must therefore be a more subtle sort, and would not be directly visible to single any freely falling observer.

In any discussion of de Sitter space, it is tempting to ask about quantum gravity effects.  Although gravitons are massless, they admit a free Euclidean vacuum state \cite{Allen:1986tt}. It is therefore plausible that our results may generalize to graviton $n$-point functions (though there is a certain tension with the results of \cite{AAS,Tsamis:1992sx,Tsamis:1994ca,Giddings:2010nc}).  Such a result would again preclude perturbative instabilities in this context -- at least in cases where they are not already present at the classical level.    However, even in this case there may still be room for more subtle quantum gravity effects associated with large regions of de Sitter space (see e.g. \cite{Giddings:2010nc,NAH,ADNTV,DO,GM}) which are not instabilities per se, and which remain to be investigated in more detail.  In addition, there are clearly interesting quantum effects involving gravity coupled to scalars with very flat potentials.  This exception is allowed due to the fact that free massless scalars are already marginally unstable (at both the quantum and classical levels).  The prime example of such an interesting quantum effect is of course eternal inflation, which will occur barring the discovery of further novel phenomena, and which may have further implications for understanding quantum de Sitter space \cite{ADNTV}.

\begin{acknowledgements}
The authors thank David Berenstein, Steven B. Giddings, James B. Hartle, Renaud Parentani, and Mark Srednicki for many useful and interesting discussions of quantum field theory in both Minkowski and de Sitter space.
We also thank Cliff Burgess, Richard Holman, Louis Leblond, and Sarah Shandera for discussions related to \cite{Burgess:2010dd}.
This work was supported in part by the US National Science Foundation under grants   PHY05-55669 and PHY08-55415 and by funds from the University of California.
\end{acknowledgements}

\appendix

\section{Notation and conventions}
\label{app:notation}

\subsection{The Gamma and related function}
\label{app:gamma}

We use the following notation in the main text. The Euler Gamma
function is denoted $\G{z}$, and we use the condensed notation
\eq{
  \GGG{a_1,\,a_2,\,\dots,\,a_j}{b_1,\,b_2,\,\dots,\,b_k}
  := \frac{\G{a_1}\G{a_2}\cdots\G{a_j}}{\G{b_1}\G{b_2}\cdots\G{b_k}} .
}
We also define the Pochhammer symbol for complex $a$ and $n \in \Nat$
\eq{
  (a)_n := \GGG{a+n}{a} = (a)(a+1)\cdots(a+n-1) ;
}
thus, $(a)_n$ is simply a polynomial of $a$ of order $n$.
The digamma function $\dig{z}$ is the logarithmic derivative of
the Gamma function  $\dig{z} := \frac{\Gamma'(z)}{\G{z}} $.

\subsection{Gegenbauer functions and polynomials}
\label{app:Geg}

The Gegenbauer function of the first kind may be defined via the
hypergeometric function
\eq{ \label{eq:Geg_def}
  C_\l^\a(z) := \GGG{2\a+\l}{1+\l,\,2\a}
  \2F1{-\l}{\l+2\a}{\a+\half}{\frac{1-z}{2}} .
}
Here $\a$, $\l$, and $z$ are arbitrary complex numbers. Important
features of this function, including its analytic properties, recursion
relations, asymptotic forms, etc., are presented in \cite{Durand}.
The function's relation to representations of $SO(n)$ and related
groups is nicely described in \cite{Vilenkin91}. Here we
present only information used in the text. For $|z| > 1$ the Gegenbauer
function may be usefully rewritten \cite{Durand}
\eqn{ \label{eq:Geg_largeZ}
  C_\l^\a(z) &=&
  \GGG{\l+2\a,\,-(\l+\a)}{\a,\,-\l,\,\l+1} (2z)^{-(\l+2\a)}
  \2F1{\frac{\l+2\a}{2}}{\frac{1+\l+2\a}{2}}{\l+\a+1}{z^{-2}}
  \nn \\ & &
  + \GGG{\l+\a}{\a,\,\l+1}(2z)^\l
  \2F1{- \frac{\l}{2}}{\frac{1-\l}{2}}{-\l-\a+1}{z^{-2}} .
}
From this we see that at large $|z| \gg 1$ the Gegenbauer functions
have two asymptotic branches, namely,
\eqn{ \label{eq:Geg_largeZ_branches}
  C_\l^\a(z) &=&
  \GGG{\l+2\a,\,-(\l+\a)}{\a,\,-\l,\,\l+1} (2z)^{-(\l+2\a)}
  \left[1 + O(z^{-2}) \right] \nn \\ & &
  +
  \GGG{\l+\a}{\a,\,\l+1}(2z)^\l \left[1 + O(z^{-2}) \right] .
}
Gegenbauer functions satisfy many recurrence relations; some that
we will make use of are
\eq{ \label{eq:Geg_recurrence}
  (\l+\a) C_\l^\a(z) = \a \left[C_\l^{\a+1}(z) - C_{\l-2}^{\a+1}(z) \right],
}
\eq{
  \frac{d^n}{dz^n} C_\l^\a(z) = 2^n (\a)_n C^{\a+n}_{\l-n}(z) .
}

When $\l = L \in \Nat$ the hypergeometric series
terminates and Gegenbauer functions reduce to the Gegenbauer polynomials.
The Gegenbauer polynomials $C_L^{\a}(z)$ form a complete orthogonal basis
on the interval $z \in [-1, 1]$ with respect to the measure $(1-z^2)^{\a-1/2}$.
They have normalization
\eq{ \label{eq:Geg_norm}
  A_L^\a := \int_{-1}^{+1} dz(1-z^2)^{\a-1/2} C_L^\a(z)C_M^\a(z)
  = \frac{\pi 2^{1-2\a}}{(L+\a)}\GGG{L+2\a}{L+1,\,\a,\,\a} \delta_{L M}.
}
Gegenbauer polynomials obey the reflection formula
\eq{ \label{eq:Geg_reflection}
  C_L^\a(z) = (-1)^L C_L^\a(-z) .
}
The integral of three Gegenbauer polynomials with common degree
$\a$ is \cite{Vilenkin91}:
\eqn{ \label{eq:Geg_triple_integral}
  D(\a; L,\, M,\, N)
  &:=& \int_{-1}^{+1} dz (1-z^2)^{\a-1/2} C^\a_L(z) C^\a_M(z) C^\a_N(z) \nn \\
  &=& \frac{2^{1-2\a}\pi}{\Gamma^4(\a)}
  \GGG{J+2\a,\,J-L+\a,\,J-M+\a,\,J-N+\a}{J+\a+1,\,J-L+1,\,J-M+1,\,J-N+1} ,
}
when $J := (L+M+N)/2 \in \Nat$, and $L$, $M$, and $N$ satisfy the triangle
inequalities; otherwise $D(\a,L,M,N) = 0$.

\subsection{The function ${}_7V_6(a;b,c,d,e,f)$}
\label{app:V76}

The function ${}_7V_6(a;b,c,d,e,f)$ is an ${}_7F_6$ hypergeometric
function with unit argument and a special form of the parameters
\cite{Slater}:
\eqn{ \label{eq:V76def}
  & & {}_7V_6(a;\,b,\,c,\,d,\,e,\,f) \nn \\
  &:=&
  {}_7F_6\left[ \begin{array}{ccccccccc}
      a, & 1+\frac{a}{2}, & b, & c, & d, & e, & f, & & \\
      & & & & & & & ; & 1 \\
      \frac{a}{2}, & 1+a-b, & 1+a-c, & 1+a-d, & 1+a-e, & 1+a-f, & & & \\
    \end{array}
  \right] .
}
The series defining (\ref{eq:V76def}) converges when it's parametric ``excess''
$s = 4+4a - 2(b+c+d+e+f)$ has a real part that is greater than zero.
The series terminates when one of the parameters is a negative integer.
When the series terminates because one of ${b,c,d,e,f}$ is a negative
integer and the excess takes the value $s = 2$ the series may be summed and the result
is known as Dougall's formula:
\eq{ \label{eq:Dougalls}
  {}_7V_6(a;\,b,\,c,\,d,\,e,\,-n)
  = \frac{(1+a)_n (1+a-b-c)_n (1+a-c-d)_n (1+a-b-d)_n }
  {(1+a-b)_n (1+a-c)_n (1+a-d)_n (1+a-b-c-d)_n }
}
with $e = 1 + 2a-b-c-d + n$ and $n \in \Nat$.

There exist a large number of relations between functions of the form
${}_7V_6(a;b,c,d,e,f)$ different parameters. One
such relation of which we will make use is
\eqn{ \label{eq:V76XF1}
  & & {}_7V_6(a;\,b,\,c,\,d,\,e,\,f) \nn \\
  &=&
  \GGG{1+a-e,\,1+a-f,\,2+2a-b-c-d,\,2+2a-b-c-d-e-f}
  {1+a,\,1+a-e-f,\,2+2a-b-c-d-e,\,2+2a-b-c-d-f} \nn \\ & & \times
  {}_7V_6(1+2a-b-c-d;\,1+a-c-d,\,1+a-b-d,\,1+a-b-c,\,e,\,f) .
}
This equality is valid so long as the series on both sides
converge, i.e. that the excess of both series is greater than zero.

It is convenient to define the regularized function
\eqn{ \label{eq:V76regdef}
  & & \V76reg(a;b,c,d,e,f)  \nn \\
  &:=& \frac{{}_7V_6(a;b,c,d,e,f)}{
    \GG{\frac{a}{2},\,1+a-b,\, 1+a-c,\, 1+a-d,\, 1+a-e,\, 1+a-f}} \\
  &=& \sum_{n=0}^\infty
  \frac{(a)_n (1+a/2)_n (b)_n (c)_n (d)_n (e)_n (f)_n}
  {\GG{1+n,\,a/2 +n,\,1+a-b+n,\, \dots\, ,\, 1+a-f+n}} .
}
This series defines an entire function in all of it's parameters. Like
${}_7V_6(a;b,c,d,e,f)$ the series terminates when one of the
parameters is a negative integer. When $-a \in \Nat$, i.e. when
$a = 0,-1,-2,\dots$, the series is zero.

\section{Calculation of $\D^{\s_1}\D^{\s_2}(Z)$ }
\label{app:DeltaDelta}

In this appendix we compute the spectral representation (\ref{eq:DeltaDeltaRep})
of the product of two free Euclidean 2-point functions on the sphere
$S^D$. As discussed in the main text, this amounts to computing
(\ref{eq:rhodef}), where $\a := d/2$.  Using the definition of  the
constant $A^\a_L$ from (\ref{eq:Geg_norm}), one may check that
$\rho_{\s_1 \s_2}(L) = - \rho_{\s_1 \s_2}(-L-2\a)$ for $L \in \Nat$,
$\s_1,\s_2 \in \mathbb{C}$.
In this appendix we consider only such positive integer $L$ unless
otherwise noted.

From (\ref{eq:rhodef}) it is clear that
$\rho_{\s_1 \s_2}(L)$ will not in general be finite. Recall that near
$Z=1$ the 2-point function diverges as $\D_\s(Z) \sim (1-Z)^{1/2-\a}$,
so the integrand (\ref{eq:rhodef}) diverges near the boundary $Z \to 1$
for $\a \ge 3/2$. We handle this divergence using dimensional regularization;
i.e., we consider $\rho_{\s_1 \s_2}(L)$
as a function of the real parameter $\a$, evaluate $\rho_{\s_1 \s_2}(L)$
for $\a < 3/2$ for which the integral (\ref{eq:rhodef}) converges, and
then define $\rho_{\s_1 \s_2}(L)$ for $\a \ge 3/2$ via analytic continuation
of our final expression. The remainder of this appendix is concerned with
computing $\rho_{\s_1 \s_2}(L)$ and then presenting a number of checks
of our work.

We now turn to evaluating (\ref{eq:rhodef}) for $\rho_{\s_1 \s_2}(L)$.
We begin by defining
\eq{
  \L_{L\s} := \frac{2(L+\a)}{\l_{L\s}} = \frac{2(L+\a)}{(L-\s)(L+\s+2\a)}
  = \frac{1}{L-\s} + \frac{1}{L+\s+2\a}.
}
and inserting (\ref{eq:2pt_free_C})  twice into (\ref{eq:rhodef}). We find
\eqn{\label{rhoSum}
  \rho_{\s_1 \s_2}(L)
  &=& \frac{2\pi^{\a+1}}{\G{\a}(L+\a)}
  \frac{1}{A^\a_L}\frac{\Gamma^2(\a)}{(4\pi^{\a+1})^2} \nn \\
  & & \times
  \sum_{M=0}^\infty \sum_{N=0}^\infty \L_{M\s_1} \L_{N\s_2}
  \int_{-1}^{+1}dZ\,(1-Z^2)^{\a-1/2} C^a_L(Z)C^\a_M(Z)C^\a_N(Z) \nn \\
  &=& \frac{\G{\a}}{8\pi^{\a+1}(L+\a)}
  \frac{1}{A^\a_L}
  \sum_{M=0}^\infty \sum_{N=0}^\infty \L_{M\s_1} \L_{N\s_2}
  D(\a; L, M, N) \nn \\
  &=:&
  \frac{1}{8\pi^{\a+1}} \GGG{L+1}{\a,\,L+2\a} S_{\s_1 \s_2}(L) .
}
To get to the second line we perform the integral using
(\ref{eq:Geg_triple_integral}), and in the third line we've defined
\eq{ \label{eq:S_def}
  S_{\s_1 \s_2}(L) :=  \sum_{M,N}{}'\, \L_{M\s_1} \L_{N \s_2}
  \GGG{J+2\a,\,J-L+\a,\,J-M+\a,\,J-N+\a}
  {J+\a+1,\,J-L+1,\,J-M+1,\,J-N+1} ,
}
where, as in (\ref{eq:Geg_triple_integral}), the sum
is over all $M$ and $N$ is such that
\eq{
  J := \frac{L + M + N}{2} \in \mathbb{N}_0, \quad
  |L-M| \le N \le L+M, \quad |L-N| \le M \le L+N .
}
We can incorporate these restrictions by a change of variables:
\eq{
  G := \frac{-L+M+N}{2} = J - L,
  \quad
  K := \frac{L+M-N}{2} = J - N,
}
such that
\eq{
  M = G+K, \quad N = G+L-K, \quad J=G+L .
}
In terms of these variables $S_{\s_1 \s_2}(L)$ becomes
\eq{ \label{eq:S_def_2}
  S_{\s_1 \s_2}(L) = \sum_{G=0}^\infty \sum_{K=0}^L
  \L_{G+K, \s_1} \L_{G+L-K, \s_2}
  \GGG{K+\a,\,L-K+\a,\,G+\a,\,G+L+2\a}
  {K+1,\,L-K+1,\,G+1,\,G+L+\a+1} .
}
In the next two sections we sum over first $K$ and then $G$.

\subsection{The $K$-sum}
\label{app:Ksum}

Let us perform the sum
\eq{
  H(L;G) := \sum_{K=0}^L \L_{G+K,\s_1} \L_{G+L-K,\s_2}
  \GGG{ K+ \a ,\, L-K+\a }{ K + 1 ,\, L-K+\a }
}
by recasting it as a contour integral in
the complex $K$-plane. We do so by multiplying the summand by
\eq{
  \pi \cot(\pi K) = - \cos(\pi K) \GG{-K,\, K+1} .
}
This function has poles with unit residue at $K=0,1,\dots,L$.
Consider then the contour integral
\eq{
  I := (-1) \oint_{\infty} \frac{dK}{2\pi i}
  \cos(\pi K) \L_{G+K,\s_1}\L_{G+L-K,\s_2}
  \frac{\GG{K+\a,\, L-K+\a}}{(-K)_{L+1}} = 0 .
}
The contour is chosen to be an arc near infinity; because the
integrand behaves at large $|K| \gg 1$ like $\sim |K|^{2\a-4}$
the integral vanishes for $\a < 3/2$. By Cauchy's formula it follows
that the sum of the residues of the poles enclosed in $C$ must likewise
sum to zero. The integrand has the following simple poles:
\begin{enumerate}
\item
  $K = 0,1,\dots,L-1,L$, due to $(-K)_{L+1}$ in the denominator,
\item
  $K = -\a -n$, $n \in \mathbb{N}_0$, due to $\G{K+\a}$ in the numerator,
\item
  $K = L+\a+n$, $n \in \mathbb{N}_0$, due to $\G{L-K+\a}$ in the numerator,
\item
  $K = -G +\s_1,\, K = -G -\s_1-2\a$, due to $\L_{G+K,\s_1}$,
\item
  $K = G +L - \s_2,\, K = G +L + \s_2+2\a$, due to $\L_{G+L-K,\s_2}$ .

\end{enumerate}
We assume for simplicity that $\s_i \neq -\a + \Int$ such that
none of the above-mentioned poles overlap. There is nothing peculiar
about these configurations and we will see that our final result is
perfectly regular at these values of the $\s_i$.

Let us now turn to evaluating the residues of these poles.
\begin{enumerate}

\item Poles at $K = 0,1,\dots,L-1,L$:
  By construction the residue of these poles sum to $-H(L;G)$.

\item Poles at $K = -\a-n$:
  These poles sum to the  infinite series
  \eq{
    \cos(\pi\a) \sum_{n=0}^\infty
    \L_{G-n-\a,\s_1} \L_{G+n+L+\a,\s_2}
    \GGG{n+\a,\,n+L+2\a}{n+1,\,n+L+\a+1} .
  }

\item Poles at $K = L+\a+n$:
  These poles sum to the infinite series
  \eq{
    \cos(\pi\a) \sum_{n=0}^\infty
    \L_{G+n+L+\a,\s_1} \L_{G-n-\a,\s_2}
    \GGG{n+\a,\,n+L+2\a}{n+1,\,n+L+\a+1} .
  }

\item Poles at $K =-G +\s_1,\, K = -G -\s_1-2\a$:
  These give two terms,
  \eq{
    \frac{\pi \cos\pi\s_1}{\sin\pi(\s_1+\a)}
    \L_{2G+L-\s_1,\s_2} \GGG{G-\s_1,\,G+L-\s_1+\a}{G+L+1-\s_1,\,G+1-\s_1-\a}
    + (\s_1 \to -(\s_1+2\a)) .
    \label{eq:Ksum_4}
  }

\item Poles at $K = G+L -\s_2,\, K = G+L+\s_2+2\a$:
  These give two terms,
  \eq{
    \frac{\pi \cos\pi\s_2}{\sin\pi(\s_2+\a)}
    \L_{2G+L-\s_2,\s_1} \GGG{G-\s_2,\,G+L-\s_2+\a}{G+L+1-\s_2,\,G+1-\s_2-\a}
    + (\s_2 \to -(\s_2+2\a)) .
  }
  These two terms are just the two terms in (\ref{eq:Ksum_4}) with
  $\s_1 \leftrightarrow \s_2$.

\end{enumerate}

Combining our results we have that
\eqn{ \label{eq:H_result}
  H(L;G) &=&
  \left\{
    \frac{\pi \cos\pi\s_1}{\sin\pi(\s_1+\a)}
    \L_{2G+L-\s_1,\s_2} \GGG{G-\s_1,\,G+L-\s_1+\a}{G+L+1-\s_1,\,G+1-\s_1-\a}
    + 3 {\rm \;sym} \right\} \nn \\ & &
  + \cos(\pi\a)\sum_{n=0}^\infty
  \bigg[ (\L_{G-n-\a,\s_1} \L_{G+n+L+\a,\s_2}
  + \L_{G+n+L+\a,\s_1} \L_{G-n-\a,\s_2})
  \nn \\ & & \phantom{+ \cos(\pi\a)\sum_{n=0}^\infty \bigg[\;\;}
  \GGG{n+\a,\,n+L+2\a}{n+1,\,n+L+\a+1} \bigg].
  \label{eq:H}
}
Here $3\rm{\;sym}$ refers to the three terms obtained by letting
$\s_1 \to -(\s_1 +2\a)$, $\s_1 \leftrightarrow \s_2$, and
$\s_1 \to -(\s_2+2\a),\; \s_2 \to \s_1$.

\subsection{The $G$-sum}
\label{app:Gsum}

Having computed the sum over $K$ we have
\eq{
  S_{\s_1 \s_2}(L)
  := \sum_{G=0}^\infty \GGG{G+\a,\,G+L+2\a}{G+1,\,G+L+\a+1} H(L;G)
}
with $H(L;G)$ given in (\ref{eq:H_result}). First let us note
that the infinite series in $H(L;G)$ (see (\ref{eq:H}))
gives a vanishing contribution when summed over $G$.
The infinite series in $H(L;G)$ contributes a term proportional to
\eqn{
  & &
  \sum_{G=0}^\infty \sum_{n=0}^\infty \bigg[
  (\L_{G-n-\a,\s_1} \L_{G+n+L+\a,\s_2}
  + \L_{G+n+L+\a,\s_1} \L_{G-n-\a,\s_2}) \nn \\ & &
  \phantom{\sum_{G=0}^\infty \sum_{n=0}^\infty \bigg[}
  \GGG{G+\a,\,G+L+2\a,\,n+\a,\,n+L+2\a}{G+1,\,G+L+\a+1,\,n+1,\,n+L+\a+1}
 \bigg].
}
Consider the change of variables $G \leftrightarrow n$; under
this action the gamma functions are invariant, as is
$\L_{G+n+L+\a,\s_i}$. However, $\L_{G-n-\a,\s} = - \L_{n-G-\a,\s}$, so
in total the summand picks up a $(-1)$ under the operation. As a result
the double sum vanishes. This statement is true for all $\a \in \mathbb{R}$ and $\s_i \in \mathbb{C}$.
So we have that
\eqn{
  S_{\s_1 \s_2}(L) &=&
  \frac{\pi \cos(\pi \s_1)}{\sin\pi(\s_1+\a)} \nn
  \sum_{G=0}^\infty \Bigg\{ \L_{2G+L-\s_1,\s_2} \nn \\ & & \times
  \GGG{G+\a,\,G+L+2\a,\,G-\s_1,\,G+L-\s_1+\a}
  {G+1,\,G+L+\a+1,\,G+L+1-\s_1,\,G+1-\s_1-\a} \Bigg\} \nn \\ & &
  + 3\rm{\;sym} .
}

We can now write $S_{\s_1 \s_2}(L)$ in terms of four so-called
``very well-poised'' hypergeometric series (see \ref{app:V76}):
\eqn{
  S_{\s_1 \s_2}(L) &=&
  \frac{\pi \cos(\pi \s_1)}{\sin\pi(\s_1+\a)}
  \GGG{\a,\,-\s_1,\,L+2\a,\,L-\s_1+\a}{L+\a+1,\,L+1-\s_1,\,1-\s_1-\a}
  \L_{L-\s_1,\s_2} \nn \\ & & \times
  {}_7V_6\left[L-\s_1+\a;\,\a,\,-\s_1,\,L+2\a,\,\frac{L-\s_1-\s_2}{2},\,
  \frac{L-\s_1+\s_2+2\a}{2}\right] \nn \\ & &
  + 3\rm{\;sym} .
}
These hypergeometric series have an excess of $s = 4-4\a$, so are
only absolutely convergent for $\a \le 1$. Assuming this, we may
re-write the hypergeometric series using the transformation
(\ref{eq:V76XF1}). The result is
\eqn{
  & & S_{\s_1 \s_2}(L) =
  \frac{\pi \cos(\pi \s_1)}{2 \sin\pi(\s_1+\a)}
  \GGG{\a,\,2-2\a,\,-\s_1,\,L+2\a,\,2+L-\s_1-\a,\,
    \frac{L-\s_1-\s_2}{2},\,\frac{L-\s_1+\s_2+2\a}{2}}
  {1-\s_1-\a,\,L+\a+1,\,L+1-\s_1,\,\frac{4+L-\s_1-\s_2-4\a}{2},\,
    \frac{4+L-\s_1+\s_2-2\a}{2}} \nn \\ & & \times
  {}_7V_6\left[1+L-\s_1-\a;\,1-\a,\,1-\s_1-2\a,\,1+L,\,
    \frac{L-\s_1-\s_2}{2},\,\frac{L-\s_1+\s_2+2\a}{2}\right] \nn \\ & &
  + 3\rm{\;sym}   .
}

\subsection{Final result and checks}
\label{app:final}

In the previous section we computed the sum $S_{\s_1 \s_2}(L)$;
inserting this into (\ref{rhoSum}) yields (\ref{eq:rho}).
Let us examine the poles in $\rho_{\s_1\s_2}(L)$ as a function of
$\a$, $\s_i$, and $L$. To do so it is useful to write this expression
in terms of the regularized series $\V76reg(a;b,c,d,e,f)$ defined in
(\ref{eq:V76regdef}):
\eqn{
  & & \rho_{\s_1\s_2}(L) =
  \frac{1}{8\pi^{\a}}
  \frac{\cos(\pi \s_1)}{\sin\pi(\s_1+\a)} \nn \\ & & \times
  \GG{2-2\a,\,-\s_1,\,L+1,\,1+L-\s_1-\a,\,
    \frac{L-\s_1-\s_2}{2},\,\frac{L-\s_1+\s_2+2\a}{2}}
  \nn \\ & & \times
  \V76reg\left[1+L-\s_1-\a;\,1-\a,\,1-\s_1-2\a,\,1+L,\,
    \frac{L-\s_1-\s_2}{2},\,\frac{L-\s_1+\s_2+2\a}{2}\right] \nn \\ & &
  + 3\rm{\;sym}   . \label{eq:rho_V76reg}
}
The function $\V76reg(a;b,c,d,e,f)$ is entire in all its arguments,
so the only possible poles arise from the gamma and trigonometric functions.

In (\ref{eq:rho_V76reg}) it appears that each of the four terms in
have poles when $\a = 1/2,1,3/2,\dots$. Upon inspection however one finds
that the $\rho_{\s_1\s_2}(L)$ is regular
when $\a$ is a positive integer, and in these cases we may simplify
our expression considerably. We record here the cases of $\a =1,2$:
\eq{
  \rho_{\s_1\s_2}(L) \bigg|_{\a=1} =
  \frac{1}{16\pi (L+1)}
  \left\{ \left[ \frac{\sin\pi(\s_1+\s_2)}{\sin\pi\s_1\sin\pi\s_2}
      \dig{\frac{L-\s_1-\s_2}{2}} + (\s \; {\rm syms}) \right]
    +2\pi \right\} ,
  \label{eq:rho_D3}
}
\eqn{
  & &  \rho_{\s_1\s_2}(L) \bigg|_{\a=2} =
  \frac{-1}{64\pi^2}\GGG{L+1}{L+4} \times \nn \\
  & &
  \bigg\{
  \frac{1}{4}
  (L-\s_1-\s_2-2)(L-\s_1+\s_2+2)(L+\s_1-\s_2+2)(L+\s_1+\s_2+6)
  \nn \\
  & & \phantom{\;\;\;}
  \times \left[\left[ \frac{\sin\pi(\s_1+\s_2)}{\sin\pi\s_1\sin\pi\s_2}
      \dig{\frac{L-\s_1-\s_2}{2}} + (\s \; {\rm syms}) \right]
    +2\pi \right] \nn \\
    & &
  +\big[
    (\cot\pi\s_1)(\s_1+2)(L+3)\left[L(L+2) + \s_1(\s_1+4) - \s_2(\s_2+4)\right]
    + (\s_1 \leftrightarrow \s_2) \big] \bigg\} .\nn \\
    \label{eq:rho_D5}
}
For $\a = 1/2$ the expression (\ref{eq:rho_V76reg}) is also finite.
However, for $\a = 3/2, 5/2, \dots$ the expression diverges. The
divergences near $\a = 3/2$ and $\a = 5/2$ are given in(\ref{eq:D4div}) and (\ref{eq:D6div}).

Next let us examine the pole structure of $\rho_{\s_1 \s_2}(L)$ as
a function of the mass parameters $\s_i$. We restrict the $\s_i$ to
be ``on-shell'', i.e. to have values corresponding to positive mass-squared
(see section \ref{sec:prelim}).
Under this restriction the only possible poles in (\ref{eq:rho_V76reg})
are due to factor $1/\sin[\pi(\s_i+\a)]$ and occur
when $\s_i = -\a + n$, $n \in \Nat$.
However, in the limit where $\s_i$ takes these values one finds
that $\rho_{\s_1\s_2}(L)$ is regular. Thus, there are no poles in
$\rho_{\s_1\s_2}(L)$ as a function of $\s_i$.

Finally, let us examine the pole structure of $\rho_{\s_1\s_2}(L)$
as a function of $L$. Recall that $\rho_{\s_1\s_2}(L)$ has been
defined only for $L \in \Nat$. For these values it's clear that
$\rho_{\s_1\s_2}(L)$ is regular. However, we may use (\ref{eq:rho_V76reg})
to extend the definition of $\rho_{\s_1\s_2}(L)$ to $L \in \mathbb{C}$.
In the complex $L$ plane this expression has several possible poles
arising from the Gamma functions
\eq{
  \GG{L+1,\,1+L-\s_1-\a,\,
    \frac{L-\s_1-\s_2}{2},\,\frac{L-\s_1+\s_2+2\a}{2}}
}
and $\s_i$ permutations. For the poles at $L=-1,-2,\dots$ one may
explicitly compute the residues using
Dougall's formula (\ref{eq:Dougalls}); the residues of the four terms
in (\ref{eq:rho_V76reg}) cancel, so in fact $\rho_{\s_1\s_2}(L)$
is regular for these values of $L$. Likewise, the Gamma functions
$\Gamma(1+L-\s_1-\a)$ and permutations do not yield poles because
their poles coincide with the zero of the series $\V76reg(a;b,c,d,e,f)$
that occur when $a$ is a negative integer. The remaining Gamma
functions do indeed yield poles in $\rho_{\s_1\s_2}(L)$. We conclude
that the expression (\ref{eq:rho_V76reg}) has poles in the complex $L$ plane at
\eqn{
  L &=& \s_1 + \s_2 - 2n,  -\s_1 + \s_2 -2 \a - 2n,  \s_1 - \s_2 -2 \a - 2n,  -\s_1 - \s_2 -4 \a - 2n  . \ \ \ \
}
We may use Dougall's formula (\ref{eq:Dougalls}) to compute the residue
at these poles:
\eqn{
  \label{eq:rhopoles}
  & & {\rm Res}\left\{ \rho_{\s_1\s_2}(L)\right\}_{L=\s_1+\s_2-2n}
  =
  \frac{-1}{8\pi^\a\Gamma(\a)}
  \frac{\sin[\pi(\s_1+\s_2+\a)]}
  {\sin\pi(\s_1+\a)\sin\pi(\s_2+\a)} \nn \\ \times & &
  \GGG
  { 1-\s_1-\s_2-2\a+2n, n+\a, n-\s_1, n-\s_2, n-\s_1-\s_2-\a }
  { -\s_1-\s_2+2n, 1+n, 1+n-\a-\s_1, 1+n-\a-\s_2, 1+n-\s_1-\s_2-2\a } .
  \nn \\
}

It is important to realize that (\ref{eq:rho}) is not the
unique extension of $\rho_{\s_1\s_2}(L)$ to complex values of $L$.
For example, under the assumption that $L \in \Nat$ one may perform
several hypergeometric transformations on (\ref{eq:rho}) to derive
an alternate expression for $\rho_{\s_1\s_2}(L)$ which agrees with (\ref{eq:rho}) for $L \in \Nat$ but has a  different pole structure in the complex $L$ plane.

\section{$D=3,4$ 1-Loop corrections from 4-particle interactions}
\label{sec:4p1l}

Here we simply list the results of the calculations outlined in section \ref{sec:1loop} for diagrams shown in figure \ref{fig:4scalar} for dimensions $D=3,4$.  The key point is that the constant $\Delta^{\s_2}(1)$ is
given formally by setting $Z = 1$
in (\ref{eq:2pt_free}):
\eq{ \label{eq:2pt_Z1}
  \D^\s(1) = \ell^{2-D} \frac{\cos\pi\left(\s+\frac{d}{2}\right)}
  {2^{d+1}\pi^{(d+3)/2}}
  \GG{-\s,\,\s+d,\,\frac{1-d}{2}} .
}
This expression diverges for $d=1,3,5,\dots$ due to the factor
$\G{1-d/2}$. In these dimensions the divergence may be cancelled
by the counterterm.

For $D=3$ the expression (\ref{eq:2pt_Z1}) is finite.  As a result, following the minimal subtraction scheme (MS) we set the counterterms to zero
and compute the self-energy correction
  $\Pi_1 \big|_{D=3} = g (1+\s)\cot\pi\s / (8 \pi \ell) $ ,
which represents a shift of particle 1's mass $M_1^2 \rightarrow M_1^2 + \Pi_1$.

For $D=4-\eps$ we have the divergent expression
\eqn{
  \D^\s(1) \bigg|_{D=4-\eps}
  &=& \frac{(1+\s)(2+\s)}{8\pi^2 \ell^2}\frac{1}{\eps}  \\
  & &
  - \frac{(1+\s)(2+\s)}{16\pi^2 \ell^2}
  \left[-1+\gamma+ \pi \cot\pi\s - \ln(4\pi) + 2 \psi(3+\s) \right]
  + O(\eps) ,\nn
}
where $\gamma$ is the Euler constant and $\psi(x)$ the digamma
function. Defining the counterterms
\eq{
  \left(\delta M_1^2 \right) \bigg|_{D=4-\eps}
  = - \frac{g}{2}\frac{(1+\s_2)(2+\s_2)}{8\pi^2 \ell^2}\frac{1}{\eps} ,
  \quad
  \left(\delta M_2^2 \right) \bigg|_{D=4-\eps}
  = - \frac{g}{2}\frac{(1+\s_1)(2+\s_1)}{8\pi^2 \ell^2}\frac{1}{\eps}
}
leads to the self-energy correction
\eq{
  \Pi \bigg|_{D=4-\eps} =
  - \frac{g M_2^2}{32\pi^2}
  \left[-1+\gamma+\log\left(\frac{M_2^2}{4\pi} \right)\right]
  + O(\eps)
}
to the $M_1^2$.  As noted in section (\ref{sec:1loop}), these mass
shifts encode the full context of the 4-particle 1-loop corrections.
Both of these expressions agree with the flat-space result
\eq{
  \Pi\bigg|_{\rm flat}
  = \frac{g}{2(4\pi)^{D/2}}\frac{\G{1-\frac{D}{2}}}{(M_2)^{2-D}} .
}



\end{document}